\documentclass[twoside,11pt]{article}
\usepackage{graphicx}
\usepackage{multicol}

\def\baselinestretch{0.95}
\font\tenrsfs=rsfs10
\font\sevenrsfs=rsfs7
\font\fiversfs=rsfs5
\newfam\rsfsfam
\textfont\rsfsfam=\tenrsfs
\scriptfont\rsfsfam=\sevenrsfs
\scriptscriptfont\rsfsfam=\fiversfs
\def\mathscr#1{{\fam\rsfsfam\relax#1}}
\def\Lag{\mathscr{L}}
\newcommand{\mb}[1]{\mbox{\normalsize\boldmath $#1$}}
\newcommand{\fig}[1]{~{\rm \ref{fig:#1}}}

\def\circa#1{\,\raise.3ex\hbox{$#1$\kern-.75em\lower1ex\hbox{$\sim$}}\,}
\def\Red  {}

\def\Black{}
\def\Blue {}
\newcommand{\eV}{\,{\rm eV}}
\newcommand{\meV}{\,{\rm meV}}
\newcommand{\keV}{\,{\rm keV}}
\newcommand{\MeV}{\,{\rm MeV}}
\newcommand{\eq}[1]{~(\ref{eq:#1})} 

\def\be{\begin{equation}}
\def\ee{\end{equation}}
\def\bc{\begin{center}}
\def\ec{\end{center}}
\def\bea{\begin{eqnarray}}
\def\eea{\end{eqnarray}}

\makeatletter
\def\art{\@ifnextchar[{\eart}{\oart}}
\def\eart[#1]#2#3#4#5#6{{\rm #2}, {\em #3 \rm #4} {\rm (#6) #5 ({\em #1})}}
\def\hepart[#1]#2{{\rm #2, \em#1}}
\newcommand{\oart}[5]{{\rm #1}, {\em #2 \rm #3} {\rm (#5) #4}}

\newcommand{\NP}{Nucl. Phys.}

\newcommand{\PRL}{Phys. Rev. Lett.}
\newcommand{\PL}{Phys. Lett.}
\newcommand{\PR}{Phys. Rev.}

\begin{document}
 \centerline{hep-ph/0201291 \hfill CERN--TH/2002--13\hfill IFUP--TH/2002--1}
\vspace{5mm}
\Black
\vspace{0.5cm}
\centerline{\LARGE\bf\Red Neutrino oscillations 
and signals in $\beta$ and
$0\nu2\beta$ experiments\footnote{In the addendum at pages \pageref{in}--\pageref{out}
we update our results including the first KamLAND data.
In the addendum at pages \pageref{insalt}--\pageref{outsalt}
we update our results including the `salt' SNO data.}}
 \medskip\bigskip\Black
  \centerline{\large\bf Ferruccio Feruglio}\vspace{0.2cm}
   \centerline{\em Universit\`a di Padova and INFN, sezione di Padova, Italia}\vspace{0.4cm}
   \centerline{\large\bf Alessandro Strumia$^\dagger$}\vspace{0.2cm}
   \centerline{\em Theoretical Physics Division,
 CERN, CH-1211 Gen\`eve 23, Suisse}\vspace{0.4cm}
   \centerline{\large\bf  Francesco Vissani}\vspace{0.2cm}
   \centerline{\em INFN, Laboratori Nazionali del Gran Sasso,
 Theory Group, I-67010 Assergi (AQ), Italy}
\vspace{1cm}
\Blue\centerline{\large\bf Abstract}
\begin{quote}\large\indent
Assuming Majorana neutrinos,
we infer from oscillation data
the expected values of the parameters
$m_{\nu_e}$ and $m_{ee}$ probed by $\beta$ and $0\nu2\beta$-decay
experiments.
If neutrinos have a `normal hierarchy'
 we get the $90\%$ CL range
$|m_{ee}| = (0.7\div 4.6)\meV$,
and discuss in which cases future experiments can test this possibility.
For `inverse hierarchy', we get
$|m_{ee}| = (12\div 57)\meV$ and
$m_{\nu_e} = (40\div 57)\meV$.
The $0\nu 2\beta$ data imply that almost degenerate neutrinos
are lighter than $1.05\,h\eV$ at 90\% CL
($h\sim1$ parameterizes nuclear uncertainties),
competitive with the $\beta$-decay bound.
We critically reanalyse the data  that were recently
used to claim an evidence for $0\nu 2\beta$, 
and discuss their implications.
Finally,
we review the predictions of flavour models
for $m_{ee}$ and $\theta_{13}$.
\Black
\end{quote}
\vspace{5mm}

\renewcommand{\thefootnote}{\fnsymbol{footnote}}
\footnotetext[2]{On leave from dipartimento di Fisica
dell'Universit\`a di Pisa and INFN.}
\renewcommand{\thefootnote}{\arabic{footnote}}

\noindent
The most successful extensions of the SM to date suggest
that neutrinos are Majorana particles, with masses around
$v^2/M_{\rm GUT}\sim\ \mbox{meV}$, where
$v=174$ GeV (the electroweak scale) and
$M_{\rm GUT}\sim 2\cdot 10^{16}$ GeV (the
grand unification scale).
If this indication is correct,
the goal of
future neutrino experiments will be the reconstruction of
 the neutrino Majorana mass matrix,
namely the measurement of
3 masses, 3 mixing angles and 3 CP-violating phases.
Within this framework, data on
solar and atmospheric neutrinos~\cite{atmexp,sunexp,reactorexp} are interpreted
as pieces of evidence for effects due to
2 mixing angles and 2 squared mass differences, and
the LSND anomaly~\cite{LSND} cannot be explained.
We are far from knowing all 9 neutrino parameters.
Furthermore, only 6 of them
might be measured by oscillations; thus, we will eventually
need ``non-oscillation'' experiments
to access the remaining ones.
In this work, we discuss how future study of
neutrinoless double beta decay ($0\nu2\beta$) 
and tritium
end-point  spectrum  ($\beta$) 
could contribute to this goal.
In particular, we carefully study the expected signals
for these experiments on the basis of our present knowledge.
We follow two different strategies of investigation:
we first pursue a purely phenomenological approach,
and next we add speculative ingredients
from models of quark and lepton masses.
In our view, neither of these approaches is entirely satisfactory:
the first one allows us to derive safe but not extremely strong restrictions;
the second one can give stronger but unsafe restrictions.
Therefore these two approaches are ``complementary''.


In section~\ref{osc} we obtain the ranges of
parameters measured by oscillation experiments,
and summarize how future experiments are expected to improve
on these quantities.
In section~\ref{nosc} we work out the connections
between oscillation and non-oscillation ($\beta$ and
$0\nu2\beta$-decay) experiments.
We improve on earlier similar works,
by combining these considerations with the
existing oscillation data by means of appropriate statistical techniques.
In this way, we determine the precise ranges of
$\beta$ and $0\nu2\beta$-decay signals expected on the basis
of present oscillation data, for allowed
neutrino mass spectra.
In the case of almost degenerate neutrinos,
we also compute the upper bound on the common Majorana neutrino mass
implied by present $0\nu2 \beta$ experiments.
An evidence for $0\nu2\beta$ has been claimed in a recent paper~\cite{evid}.
In the appendix we critically reanalyse the data 
that should contain such evidence
and discuss their implications.
In section~\ref{models} we review the expected values of $m_{ee}$ and
$\theta_{13}$ (the two
still unknown parameters with brighter
future experimental prospects)
in large families of flavour
models discussed in the literature.
Our results are summarized in section~\ref{conclusions}.

\section{Oscillation experiments}\label{osc}
The three-flavour Majorana neutrino mass matrix
$m_{\ell \ell'}$ (where $\ell,\ell'=\{e,\mu,\tau\}$)
is described by 9 real parameters.
It is convenient to choose them to be
the 3 real positive masses $m_i$
and parameterize the neutrino mixing matrix $V$
(that relates the fields with given flavour
to those with given mass, $\nu_\ell=V_{\ell i} \nu_i$) as
\begin{equation} V =
R_{23}(\theta_{23}) \cdot
\hbox{diag}\,(1, e^{i \phi},1) \cdot
R_{13}(\theta_{13}) \cdot
R_{12}(\theta_{12}) \cdot
\hbox{diag}\,(1 , e^{i \alpha}, e^{i \beta})
\label{eq:Vunitary}
\end{equation}
where $R_{ij}(\theta_{ij})$ represents a
rotation by $\theta_{ij}$ in the $ij$ plane
and $i,j=\{1,2,3\}$.
The CP-violating phase $\phi$ induces no physical effects if
$\theta_{13}=0$.
Similarly, also $\alpha$ and $\beta$ induce physical effects only in presence
of flavour mixing, but we found no simple way
of presenting the necessary conditions
in terms of the above parametrization.

We order the neutrino masses $m_i$ such that
$m_3$ is the most splitted state and
$m_2 > m_1$, and define
$\Delta m^2_{ij}=m_j^2-m_i^2$.
With this choice, $\Delta m^2_{23}$
and $\theta_{23}$ are the `atmospheric parameters' and
 $\Delta m^2_{12}>0$ and $\theta_{12}$ are the `solar parameters',
whatever the spectrum of neutrinos
(`normal hierarchy' i.e.\ $m_1\ll m_2\ll m_3$ so that
$\Delta m^2_{23}>0$;
`inverted hierarchy'  i.e.\ $m_3\ll m_1\circa{<} m_2$, so that $\Delta
m^2_{23}<0$ or `almost degenerate').

\subsection{Present situation}
\begin{figure}[t]
$$\hspace{-8mm}\includegraphics[width=188mm]{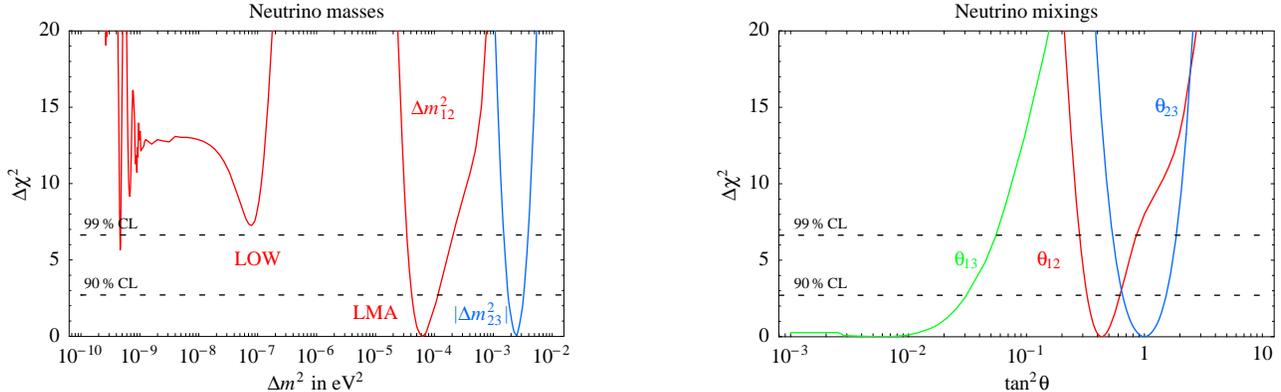} $$
  \caption[]{\em Summary of present
data on neutrino masses
and mixings.
\label{fig:panta}}
\end{figure}

Almost all present experimental information
on the 9 neutrino parameters
comes
from oscillation experiments~\cite{atmexp,sunexp,reactorexp},
 and can be summarized as
\begin{equation}
|\Delta m_{23}^2|^{1/2}\sim 50\ \mbox{meV},\quad
(\Delta m_{21}^2)^{1/2}\sim (0.03\div 25)\ \mbox{meV},\quad
\theta_{12}\sim \theta_{23}\sim 1,\quad
\theta_{13}\circa{<} 1/4,\quad \phi \sim (0\div 2\pi)
\label{osciPara}
\end{equation}
and more precisely in fig.~\ref{fig:panta},
obtained from our up-to date fits of solar, atmospheric and reactor data.
Details on how our fits have been performed can be found in~\cite{sunfit,atmfit}.
The atmospheric mixing angle $\theta_{23}$
and the solar mixing angle $\theta_{12}$
have to be large (indeed, it looks as
if the first is almost maximal, while the second
may be just `large');
the {\sc Chooz} experiment
implies that the third mixing $\theta_{13}$ cannot be that large.
The solar mass splitting, $\Delta m^2_{12}$, could be slightly or much
smaller than the atmospheric one, $|\Delta m^2_{23}|$.
Nothing is known on $\phi$.

\medskip

A brief comment about statistics could help in better interpreting
our results.
In fig.~\ref{fig:panta} and in the rest of the paper
we plot $\Delta \chi^2(p)\equiv \chi^2(p) - \chi^2_{\rm best~fit}$,
where $p$ is the quantity in which
we are interested;
$\chi^2(p)$ is usually obtained from a multi-parameter fit, as
$ \chi^2(p)= \min_{p'}\chi^2(p,p')$,
where $p'$ are other parameters in which we are not interested.
For example, a fit of atmospheric data gives
$\chi^2(\Delta m^2_{23},\theta_{23},\theta_{13})$,
from which we obtain the functions
$\Delta\chi^2(\Delta m^2_{23})$ and $\Delta\chi^2(\theta_{23})$,
plotted in figs.~\ref{fig:panta}a and \ref{fig:panta}b respectively.
It is important to appreciate that, when
a quantity is extracted in this way from a multi-parameter fit,
{\em there is no loss of statistical power due to the presence of
other parameters}.
In fact, with some technical assumptions (e.g.\ the
Gaussian approximation\footnote{When extracting the
oscillation parameters from present data,
only in one case this assumption is not sufficiently well realized in practice.
For simple physical reasons
the Beryllium contribution to solar neutrino rates
depends in very strong oscillatory way on $\Delta
m^2_{12}$, for $\Delta m^2_{12}\sim 10^{-9}-10^{-10}\eV^2$
(see fig.~\ref{fig:panta}a).
Therefore the $\Delta\chi^2(\theta_{12})$ shown in
fig.~\ref{fig:panta}b has been
obtained  from a full Bayesian analysis, performed assuming a flat prior
in $\ln\Delta m^2_{12}$ along the lines of~\cite{sunfit,GG}.
Only in this case, and only around $\theta_{12}\sim 1$,
there is some difference between a Bayesian
result and the Gaussian approximation,
and we use the Bayesian result.
})
one can derive the following two equivalent results:
(1) In the frequentistic framework
 $\Delta\chi^2(p)$ is
distributed as a $\chi^2$ with {\em one} degree of freedom.
(2) In the Bayesian framework
$\exp[-\Delta\chi^2(p)/2]$ is the probability
of different $p$ values,
up to a normalization factor.
Our inferences on $m_{ee}$ also depend on unknown parameters
($\theta_{13}$ and the CP-violating phases):
using the Gaussian approximation
we obtain more simple and conservative results,
as explained in section~\ref{How?}.

\medskip

In fig.\fig{panta} and in the rest of the paper
we do not include the significant but controversial
information from SN1987A,
that would disfavour $\theta_{13}\circa{>}1^\circ$ (if $\Delta m^2_{23}<0$)
{\em and} solar solutions with large mixing angle~\cite{SNatm,SNsun}.
However, we recall here the origin of these bounds.
The average $\bar{\nu}_e$ energy deduced from Kamiokande
II and IMB data is $E_{\bar{\nu}_e} \sim 11\MeV$, assuming
the overall flux suggested
by supernova simulations (experimental data alone do not
allow to extract both quantities accurately).
This is somehow smaller than the value suggested by
supernova simulations in absence of oscillations,
$E_{\bar{\nu}_e} \sim 15\MeV$. For both figures
it is difficult to properly assign errors;
but oscillations that convert $\bar\nu_e \leftrightarrow
\bar{\nu}_{\mu,\tau}$ increase the disagreement,
since supernova simulations suggest
$E_{\bar{\nu}_{\mu,\tau}}\sim 25\MeV$.
With an inverted hierarchy, $\theta_{13}\circa{>}1^\circ$
gives rise to adiabatic MSW conversion,
swapping $\bar{\nu}_e \leftrightarrow
\bar{\nu}_{\mu,\tau}$ completely.
This is why
this case is `disfavoured'
if the predictions of supernova models
on neutrino energy and flux are correct.
The same argument applies to
large solar mixing angles:  $\theta_{12}\sim 1$
induces a partial swap of the $\bar{\nu}_e$
into $\bar{\nu}_{\mu,\tau}$, whatever the mass spectrum of neutrinos.
LMA oscillations have a smaller $\theta_{12}$
and a larger $\Delta m^2_{12}$ than LOW and (Q)VO,
and are therefore less `disfavoured'. SMA gives almost
no $\bar{\nu}_e$ oscillations, but is strongly disfavoured
by solar data. For a full analysis, see~\cite{SNsun}.

\subsection{Perspectives of improvement}
Future oscillation experiments can significantly improve the situation.
Concerning the `solar' parameters,
SNO, KamLAND and Borexino can reduce the
error on $\sin^22\theta_{12}$ down to around $5\%$,
and measure $\Delta m^2_{12}$ to few
per-mille (if it lies in the VO or QVO regions), or
few per-cent (in the LMA region), or
around $10\%$ (in the LOW
region)~\cite{jhep}.\footnote{If $\Delta m^2_{12}\circa{>} 2~10^{-4}$ a new
reactor experiment
with a shorter baseline than KamLAND would be necessary~\cite{jhep,PetcovReactor}.
If $\Delta m^2_{12}\approx 10^{-8}\eV^2$ Borexino and KamLAND
will not see a inequivocable signal.}
Concerning the `atmospheric' parameters,
K2K, Minos or CNGS can reduce the error on
$|\Delta m^2_{23}|$ and $\sin^2 2\theta_{23}$ down to about $10\%$
and discover $\theta_{13}$ if larger than few degrees~\cite{Para,icarus}
(the precise value strongly depends on $|\Delta m^2_{23}|$).
Far future long-baseline experiments can reduce the error on
$|\Delta m^2_{23}|$ and $\sin^2 2\theta_{23}$ down to few $\%$
(with a conventional beam~\cite{jhf}) and maybe $1\%$
(with a neutrino factory beam~\cite{nuf}).
These experiments could also discover
a $\theta_{13}$ larger than $0.5^\circ$
and tell something about $\phi$,
if LMA is the true solution of the
solar neutrino problem.
Future reactor experiments~\cite{reactor} can be sensitive to a $\theta_{13}\circa{>} 3^\circ$.

If a non zero $\theta_{13}$ will be discovered,
earth matter corrections to  $\nu_\mu\to\nu_e$
and $\bar{\nu}_\mu\to\bar{\nu}_e$
will tell the sign of $\Delta m^2_{23}$~\cite{Lipari} (i.e.\
if the atmospheric
anomaly is due to the lightest or heaviest neutrinos).
The sign of $\theta_{23}-45^\circ$ (which tells
whether the neutrino state with mass $m_3$
contains more $\nu_\tau$ or more $\nu_\mu$)
can be measured by comparing
$$P(\nu_e\to \nu_e) = 1- \sin^2 2\theta_{13} \sin^2 \frac{\Delta m^2_{23}
L}{4E_\nu}
\qquad\hbox{with}\qquad
P(\nu_\mu\to \nu_e) = \sin^2 \theta_{23}
\cdot [1-P(\nu_e\to \nu_e)]$$
(even including matter effects in $P(\nu_\mu\to \nu_e)$, the issue 
on $\theta_{23}$ remains). 
Note that muon disappearance experiments alone cannot
distinguish $\theta_{23}$ from $90^\circ-\theta_{23}$,
and that the present bound $\sin^2 2\theta_{23}\circa{>} 0.95$~\cite{atmexp}
allows the relatively loose range
$1/3\circa{<}\sin^2\theta_{23}\circa{<}2/3$.
If $\Delta m^2_{12}$ is in the upper part of the LMA region,
so that it affects long-baseline experiments, the sign of
$\theta_{23}-45^\circ$ can be measured even if $\theta_{13}=0$.

We can summarize the content of this section by saying that
we can confidently include among the known parameters
$|\Delta m^2_{23}|$, $\theta_{23}$, $\theta_{12}$,
and we have reasonably good perspectives for
$\Delta m^2_{12}$, $\theta_{13}$, ${\rm sign}(\Delta m^2_{23})$,
$\theta_{23}-45^\circ$, $\phi$
(we express our expectations by the order
of quotation; however, effects due to $\theta_{13}$ or $\phi$
could be too small to be ever observed).

\section{Non-oscillation experiments}\label{nosc}
Oscillation experiments cannot access to
3 of the 9 parameters of the Majorana mass
matrix: the overall neutrino scale,
and the two CP-violating ``Majorana phases'' denoted
as $\alpha$ and $\beta$
in eq.~(\ref{eq:Vunitary})~\cite{petcov}.
Indeed, the neutrino mass matrix enters as $m^\dagger m$ in the 
Hamiltonian of propagation of ultrarelativistic neutrinos.
Non oscillation experiments are needed to probe the
full neutrino mass matrix~\cite{doi}.
For definiteness, we focus the discussion
on $\beta$-decay tritium experiments~\cite{tri} and
$0\nu2\beta$-decay germanium experiments~\cite{ge},
that seem to have the best chances of reaching the necessary
sensitivity.
These techniques
are among the oldest and better established.
We will not discuss different techniques that at present
are not competitive, though they have
certain advantages. For instance
calorimetric measurements~\cite{gatti}
are not limited to the end-point of $\beta$ spectrum.
Search of $0\nu 2\beta$ with
other nuclear species could allow
to reduce nuclear theoretical uncertainties~\cite{fiorini}.

Before continuing, we recall other possibilities to 
investigate the mass of neutrinos offered 
by the astrophysics and the cosmology:
\begin{itemize}
\item At the next gravitational
collapse of a {\bf supernova},
the general strategy~\cite{cei} will consist in identifying
structures in the time and/or energy distributions of
neutrinos  sensitive to neutrino masses, as the neutronization peak,
the rising (or falling) ramp of the
cooling phase, a hypothetical sharp cutoff due to
black hole formation.
The sensitivity of these approaches has been quantified
in several works, assuming the capabilities of
present detectors (SuperKamiokande, SNO, LVD,\ldots).
The difference in time of flight
between neutrinos and gravitons
will only be sensitive to neutrinos heavier than
about $1\eV$~\cite{fargion},
comparable to present $\beta$-decay bounds.
The difference in time of flight between
different neutrinos will only be sensitive to neutrino
mass {\em differences}
larger than few $10\eV$~\cite{bea1}.
If neutrino emission were suddenly
terminated by black hole formation,
a  measurement of
the difference in time of flight
between neutrinos of different energy
will only be sensitive to neutrino masses larger than few eV~\cite{bea2}.

\item
Information  on neutrino masses might also be obtained from data on
{\bf large scale structures in the universe}, together with very accurate
measurements of anisotropies in the temperature of the cosmic
background radiation.
It is thought that it is possible to improve on the present limit
$\sum_\nu m_\nu \circa{<} (5\div 10)\eV$ down to
about $ \sim0.3\eV$~\cite{cosm}
(the precise value depends on uncertain cosmological parameters).
However, even
if an effect due to neutrino masses will be detected,
it will be difficult to ascertain that it is really due to neutrino masses,
rather than to other mechanisms that could produce similar effects.
Cosmology has better sensitivity than other experiments
to heavy {\em sterile} neutrinos.

\item The `{\bf $Z$ burst}'~\cite{Zburst} is a possible source of
the ultra-high energy cosmic ray events observed above the
Greisen-Zatsepin-Kuzmin (GZK) cutoff~\cite{GZK}.
A cosmic ray neutrino that scatters with
a nonrelativistic cosmic microwave background neutrino
encounters the $Z$ resonance at the energy
$E_\nu^{\rm res} =M_Z^2/2m_\nu$, that is slightly above
the optimal value even if neutrinos are as heavy as possible:
$m_\nu = 1\eV$ gives  $E_\nu^{\rm res} = 4 \times 10^{21}$ eV.
It seems difficult to imagine
a cosmological source that produces enough energetic neutrinos without
producing, at the same time, too many photons~\cite{Zburst}.

\end{itemize}

\begin{figure}[tb]
$$
\includegraphics[width=160mm]{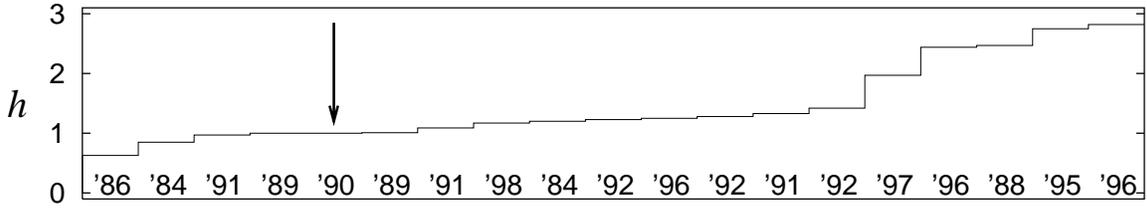}
$$
  \caption[]{\em We parameterize by $h$ the 
theoretical uncertainty on the inferences of 
$|m_{ee}|$ from the data
due to the disagreement among 
different calculations of 
nuclear matrix elements
(see eq. (\protect{\ref{nosciPara}}) 
and discussion therein).
In the horizontal scale, the year of the calculation.
Compiled from \cite{compilation}.
\label{fig:nuc}}
\end{figure}

\subsection{$\beta$-decay and $0\nu2\beta$ experiments\label{sect:noOscExps}}
Present $\beta$-decay experiments (``direct mass search'')
and neutrinoless double $\beta$-decay ($0\nu2\beta$) experiments
give bounds
on combinations of entries of
the neutrino Majorana mass matrix $m_{\ell\ell'}$.
Such bounds are somewhat above the mass scale suggested by oscillations:
at $95\%$ CL
\begin{equation}
m_{\nu_e} \equiv (m^\dagger m)_{ee}^{1/2} <2.2\ \mbox{eV}\quad
\hbox{from $\beta$-decay~\cite{mainz}}\qquad\hbox{and}\qquad
|m_{ee}|<0.38h\ \mbox{eV}\quad\hbox{from $0\nu2\beta$~\cite{HM}}.
\label{nosciPara}
\end{equation}
Here $h\equiv \mathscr{M}_0/\mathscr{M}$ is the inverse of
the nuclear matrix element $\mathscr{M}$ for $0\nu2\beta$ of
${}^{76}\mbox{Ge}$, normalized to a reference matrix
element $\mathscr{M}_0$, chosen to be the one of ref.\ \cite{m0}.
$0\nu 2 \beta$ experiments try to measure (or bound) $|m_{ee}|$ by
measuring the rate of the transition
$\Gamma(^{76}\hbox{Ge}\to ^{76}\!\hbox{Se}\,
2\beta )=\Phi\cdot \mathscr{M}^2 \cdot
|m_{ee}^2|$, where $\Phi$ is a phase space factor.
Any uncertainty on the nuclear matrix element
$\mathscr{M}$ reflects directly on the measurement of $|m_{ee}|$.
Indeed, different calculations find rather different nuclear matrix
elements, as shown in  fig.\ \ref{fig:nuc}, that suggests
the range $h=0.6\div 2.8$.
We will not attempt to assign an error on $\mathscr{M}$.
Rather, we always include an explicit factor $h$
whenever we quote an experimental result on $0\nu 2\beta$,
as we do in eq.~(\ref{nosciPara}).

The result of the Heidelberg--Moscow 
experiment at Gran Sasso (HM) 
quoted above~\cite{HM} was obtained by selecting
a certain search window, and comparing the $n=21$ events occurring
there with the $b=20.4\pm1.6$  expected background events.
Knowing  
that $m_{ee}=350\,\meV$ would yield $s=9.3/h^2$ signal events,
from the Poissonian probability of measuring $s+b$ events
one obtains the likelihood for $m_{ee}$, ${\cal L}\propto  e^{-s} (1+s/b)^n$.

\smallskip

The result of the {\sc Mainz} collaboration~\cite{mainz} quoted above
is in agreement  with {\sc Troitsk} results~\cite{troitsk},
though the $\beta$ spectrum of this last
experiment presents some unexpected features. Anyhow, this limit is
rather close to the sensitivity of these
setups, $\sim 2$ eV,
so that new experiments are certainly needed to progress.
Indeed, there are plans to
extend the sensitivities of $\beta$ and especially $0\nu2\beta$ experiments
to mass scales closer to those suggested by
oscillations, shown in eq.\ (\ref{osciPara}):
\begin{equation}
m_{\nu_e}\to 300\ \mbox{meV},\qquad
|m_{ee}|\to 10h \ \mbox{meV}.
\label{noOscFut}
\end{equation}
We refer here to the proposals of the collaborations
{\sc Katrin}~\cite{katrin} for $\beta$-decay and
{\sc Genius}~\cite{genius} for $0\nu2\beta$-decay,
that evolve from the experiments mentioned above.
There is also another proposal (named {\sc Majorana})~\cite{majorana} to 
improve on $0\nu 2\beta$ with $^{76}$Ge,
which evolves from the second most powerful experiment, 
{\sc IGEX}~\cite{igex}.
The estimated sensitivity is several times weaker than what the
{\sc Genius} experiment would like to achieve. 
There is also another interesting proposal 
named {\sc GEM}~\cite{gem}.

One should note that in eq.s~(\ref{nosciPara}) and~(\ref{noOscFut})
we suggest that $\beta$-decay experiments will at best extract
a single parameter,
$m_{\nu_e}.$
This is the case of the three neutrino scenario considered here,
for the mass splittings suggested by oscillations
are too little to be resolved by any experiment
till now proposed.\footnote{We approximate
the exact formula for the $\beta$-decay spectrum
close to end-point in presence of mixed neutrinos
in terms of a single effective neutrino
mass parameter \cite{mainz}
$m_{\nu_e}^2\equiv  |V_{ei}^2| m_i^2$ as
$$\frac{dN}{dE_\nu} \propto\sum_i |V_{ei}^2|\, \Gamma(m_i)\approx
\Gamma(m_{\nu_e})\qquad\hbox{where}\qquad
\Gamma(m)\equiv E_\nu\ \hbox{Re}\sqrt{E_\nu^2 - m_i^2} . $$
(the measured electron energy in the $^3{\rm H}\to  ^3{\rm He}\,
e\, \bar{\nu}_e$, decay is related to $E_\nu$ by kinematics).
This approximation is trivially a good one if neutrinos
are almost degenerate. However, its usefulness is more
general \cite{now}, since:
{\em (1)}~the difference between the approximated and exact spectrum,
integrated around the end-point,
vanishes at order ${\cal O}(m_i^2)$
(this is interesting for end-point search
of neutrino mass with limited energy resolution);
{\em (2)}~far from the end-point, the difference is zero at
order ${\cal O}(m_i^2)$
(this is interesting for calorimetric search of neutrino mass).
These properties do not hold for other definitions of the `effective mass',
say $m_{\nu_e}'=|V_{ei}|^2\cdot m_i$ \cite{beta}. However,
if a future $\beta$ decay experiment will attain a
very high sensitivity, and at the same time will be
able to resolve the separation between the mass levels, it
will be necessary to introduce
more parameters to describe the measured $\beta$ spectrum,
and it will be possible to extract more interesting information.
}
For example the energy resolution of {\sc Katrin}
will be more than one order of magnitude
larger than the scales in eq.~(\ref{osciPara})~\cite{badl}.
Only if neutrinos have an almost degenerate spectrum
the $\beta$-decay experiments mentioned here
could see neutrino masses.
We will consider this case closely
in section~\ref{sect:d} (for a discussion
of related matter, and more general possibilities,
see \cite{now,beta}).

\subsection{How to use the inputs from oscillations?}\label{How?}

The connection of oscillations with $\beta$ and $0\nu2\beta$-decay has
been explored in a number of
works~\cite{0nu2betaA,mee,0nu2betaB,now,beta}.\footnote{It is commonly
assumed that the dominant contribution to $0\nu 2 \beta$ comes from massive neutrinos.
We remark that, even with this assumption
there might be surprises: e.g.\ if CPT is violated,
the rate  of double $\beta$-transition
could be different from that of
double $\beta^+$, or the
$\beta^-$ absorption (EC) followed by $\beta^+$ emission.
However, we will not consider these possibilities.}
The quantities probed by $\beta$ and $0\nu2\beta$ experiments
can be written in terms
of the masses $m_i$, of the mixing angles $\theta_{ij}$ and
of the CP-violating phases $\alpha$ and $\beta$ as
\begin{eqnarray}\label{eq:mnue}
m_{\nu_e} &=&
\bigg(\sum_i |V_{ei}^2|\ m_i^2\bigg)^{1/2} =
\bigg(\cos^2 \theta_{13}(m_1^2 \cos^2\theta_{12} +
m_2^2  \sin^2\theta_{12}) + m_3^2 \sin^2\theta_{13}\bigg)^{1/2}\\
|m_{ee}| &=& \bigg|\sum_i V_{ei}^2\ m_i \bigg| =
\bigg|\cos^2 \theta_{13}(m_1
\cos^2\theta_{12} +  m_2 e^{2i\alpha} \sin^2\theta_{12}) + m_3 e^{2i\beta}
\sin^2\theta_{13}\bigg|.\label{eq:02}
\end{eqnarray}
In both formul\ae, we can identify the 3 individual contributions
associated with the 3 masses $m_i$.
It is useful to make few general remarks:
\begin{enumerate}\label{remarks}
\item $m_{\nu_e}$ depends on oscillation parameters
and on the overall neutrino mass scale,
while $m_{ee}$ is also sensitive to the Majorana phases.

\item
The $m_2$-contributions to $m_{\nu_e}$ and $m_{ee}$ are guaranteed
to be non-zero (for inverted spectrum the same is true for the $m_1$-contribution).

\item Since $|V_{ei}|\le 1$, the mixing factors suppress more strongly
the $m_i$-contributions to $m_{ee}$ than those to
$m_{\nu_e}$.
For example, at the best-fit LMA solution $V_{e2}^2\approx 1/3$
we have a $m_2$-contribution to
$m_{\nu_e}$ $70\%$ larger than the one to $m_{ee}$.
This is even more evident is the hierarchical case, 
when the $m_3$-contributions are suppressed by 
the small angle $\theta_{13}$.

\item
Let us denote as $m_{\rm min}$ the lightest neutrino mass:
$m_{\rm min}=m_1$ $(m_{\rm min}=m_3)$ in the case of normal (inverted) hierarchy.
Increasing $m_{\rm min}^2$
increases
$m_{\nu_e}^2$ by the same amount,
and $m_{\nu_e}$ is always larger than $m_{\rm min}$.
%
Instead, the behaviour of
$m_{ee}$ as a function of $m_{\rm min}$ is less simple,
especially when the individual
contributions become comparable in size.

\item Most importantly,
while the contributions to $m_{\nu_e}$  are all positive,
the individual contributions to $m_{ee}$ may
compensate each other for certain values of the
Majorana phases.
\end{enumerate}
Let us elaborate on the last point, enlarging the theoretical perspective.
One should realize that {\em we cannot even invoke
`naturalness' to argue that $m_{ee}$ should not be much smaller than its
individual $m_i$-contributions} \cite{wolfs}.
In fact, from a top/down point of view the
$ee$ element of the neutrino mass matrix is a very
simple object: some
flavour symmetry could easily force a small $m_{ee}$,
giving rise to {\em apparently} unnatural cancellations
when $m_{ee}$ is written in terms of the low energy parameters
$m_i$ and $\theta_{ij}$, as in eq.~(\ref{eq:02}).\footnote{An 
analogous remark holds in supersymmetric
see-saw models, that generate $\mu\to e \gamma$ (and other) decay amplitudes
approximately proportional to 
the $e\mu$ element of the squared neutrino Yukawa matrix.
Even in this case a flavour symmetry that demands a small $e\mu$ element
would give rise to apparently unnatural cancellations,
when $\mu\to e \gamma$ is written in terms of low energy parameters.}
A small $m_{ee}$ is stable under quantum corrections,
that renormalize it multiplicatively.
For these reasons,
bottom/up oscillation analyses that try to
determine the range of $m_{ee}$ consistent with
oscillation data must consider
seriously the most {\em pessimistic} case.
As discussed in~\cite{Bayes} naturalness considerations
would be automatically incorporated in a Bayesian analysis
(i.e.\ no need to invent `fine-tuning parameters')
if we assumed any almost flat prior probability distribution
for the phenomenological parameters
(the neutrino masses, mixing angles, and CP-violating phases).
This is why in this work we stick to the Gaussian approximation.

\smallskip

The expression of $|m_{ee}|$
minimized in the Majorana phases $\alpha$ and $\beta$
was given in~\cite{mee}. In this work, we will use it,
together with the inputs from oscillations, to infer 
the signals expected in $0\nu 2\beta$ and $\beta$-decay
experiments in a number of interesting limiting cases, namely:
normal hierarchy, inverted hierarchy, almost degenerate neutrinos.
In the first two cases (sections \ref{sect:n} and \ref{sect:i}), 
we will further show the expression of $|m_{ee}|$ which results from 
further minimization in the lightest neutrino mass $m_{\rm min}$;
this is useful to discuss why (with present data) 
we cannot exclude the possibility $m_{ee}=0$, even 
if neutrinos have Majorana 
masses and oscillate.

\begin{figure}[t]
$$\hspace{-4mm}
\includegraphics[width=8cm]{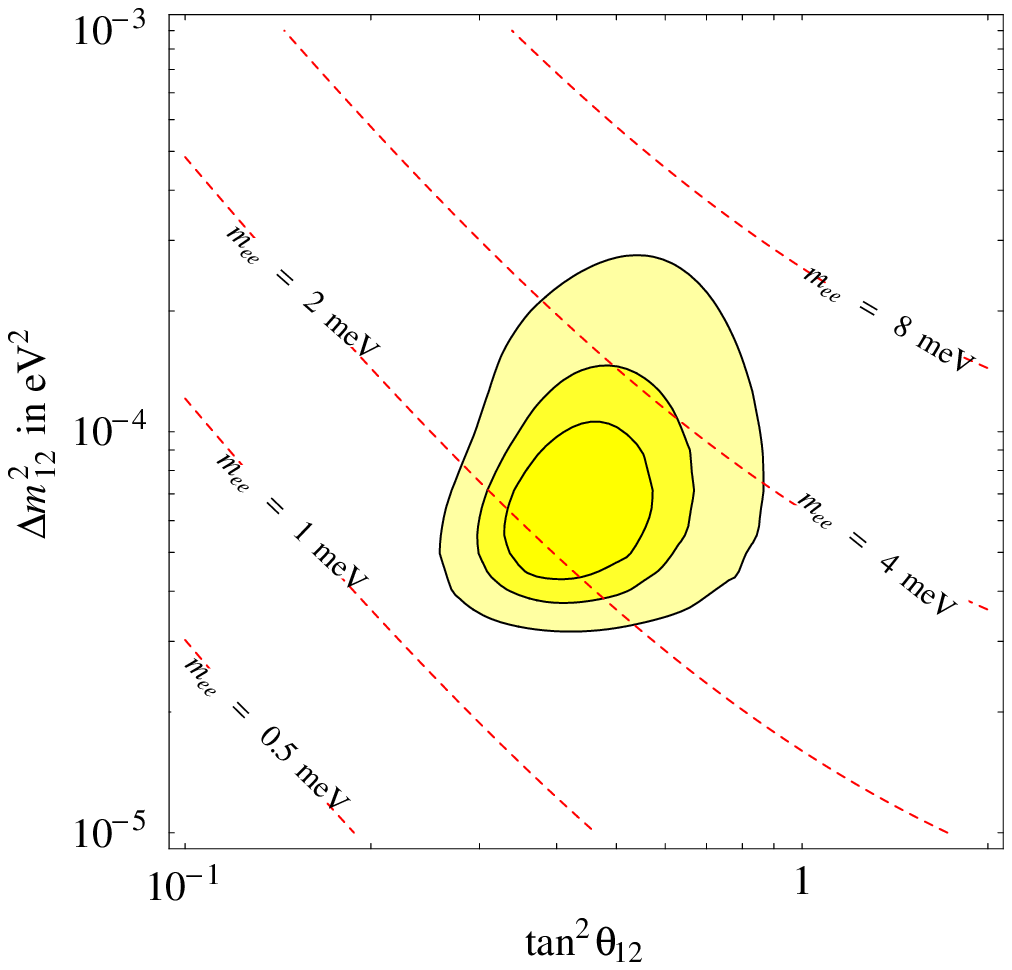}\hspace{8mm}
\includegraphics[width=8cm]{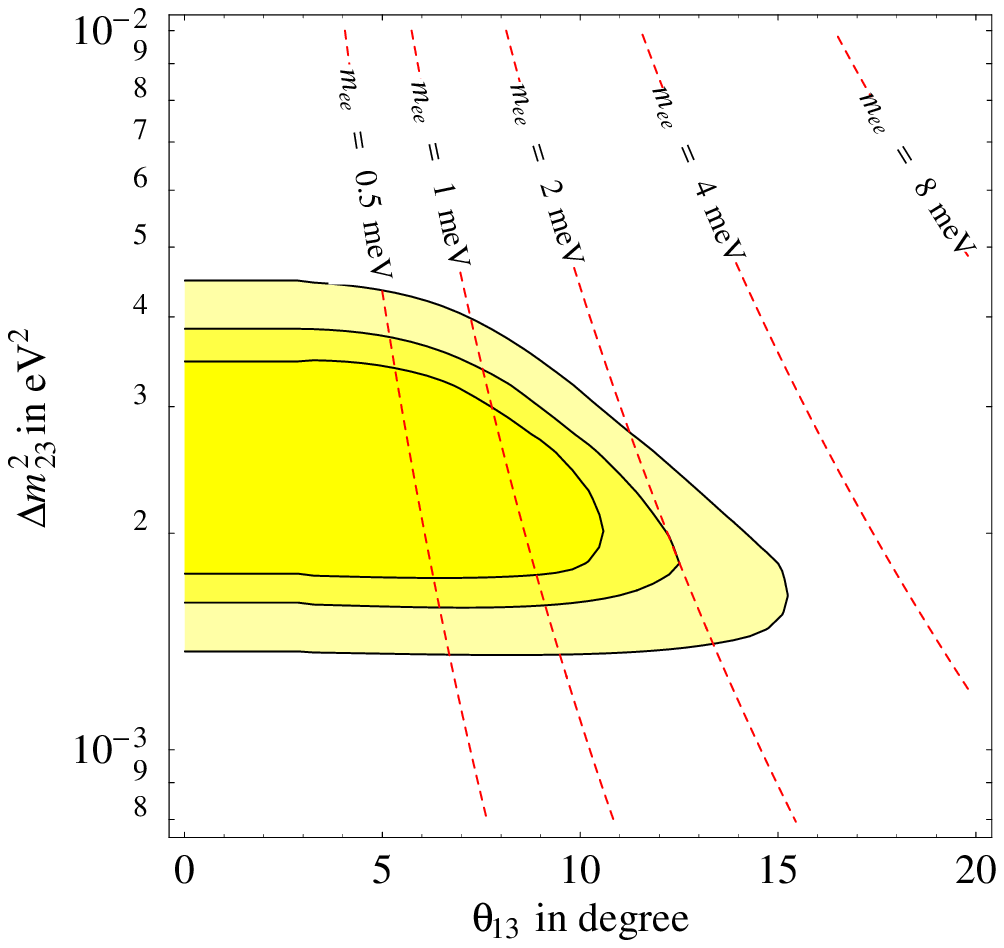}
$$
  \caption[]{\em Contour lines of the `solar' and `atmospheric' contributions
to $m_{ee}$ in the case of normal hierarchy ($m_1\ll m_2\ll m_3$),
superimposed to the regions allowed at $68,90,99\%$ CL by oscillation data.
\label{fig:iso}}
\end{figure}

\subsection{Normal hierarchy}\label{sect:n}
If neutrino masses have a partial hierarchy
$m_1\circa{<}m_2\approx (\Delta m^2_{12})^{1/2} \ll m_3\approx(\Delta m^2_{23})^{1/2}$
we could be sure that  $m_{ee}$ is non-zero:
\begin{equation}\label{eq:nger}
|m_{ee}|>
(-\Delta m^2_{12} \cos 2 \theta_{12})^{1/2} - \theta_{13}^2 (\Delta
m^2_{23})^{1/2}
\label{sure1}
\end{equation}
(up to higher orders in
$\theta_{13}$ and ${\Delta m^2_{12}}/{\Delta m^2_{23}}$,
and barring other possibilities already
excluded by oscillation data) only if the following conditions hold
\begin{equation}\label{eq:normalif}
\theta_{12} > \frac{\pi}{4}\qquad\hbox{and}\qquad
\theta_{13} < \bigg[-
\frac{\Delta m^2_{12}}{\Delta m^2_{23}}\cos 2\theta_{12}\bigg]^{1/4} .
\end{equation}
Thus,
there is no warranty to have a detectable value of
$m_{ee}$, since solar data allow $\theta_{12} > {\pi}/{4}$
at a reasonable confidence level only in the LOW and (Q)VO
solutions, where $\Delta m^2_{12}$ is too small
to give a detectable contribution to $m_{ee}$.
The minimal guaranteed $m_{ee}$ is always smaller than
the `solar' mass scale $(\Delta m^2_{12})^{1/2}$.

\smallskip

We now specialize our discussion to the case of `normal hierarchy' $m_1\ll m_2 \ll m_3$.
As discussed in section \ref{sect:nth} this 
case naturally arises in see-saw models,
even in presence of large mixing angles. A negligible $m_1$ is a prediction
of those see-saw models where only 2 right-handed neutrinos give
a significant contribution to $m_{\ell\ell'}$.

\medskip

\begin{figure}[t]
$$\hspace{-4mm}
\includegraphics[height=5.5cm]{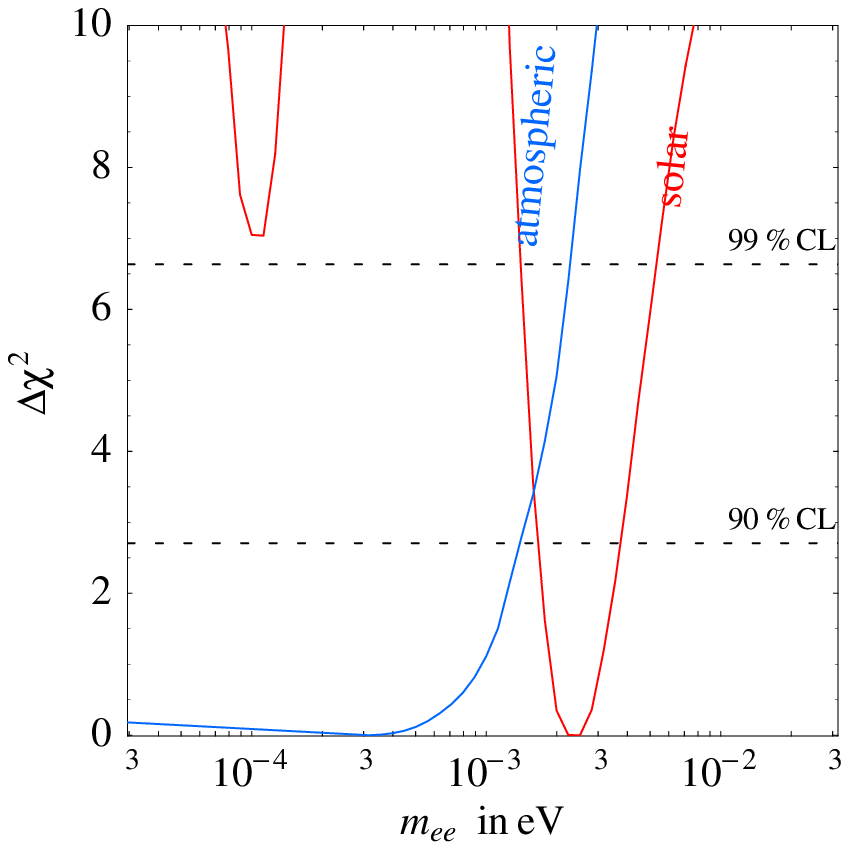}\hspace{1cm}
\includegraphics[height=5.5cm]{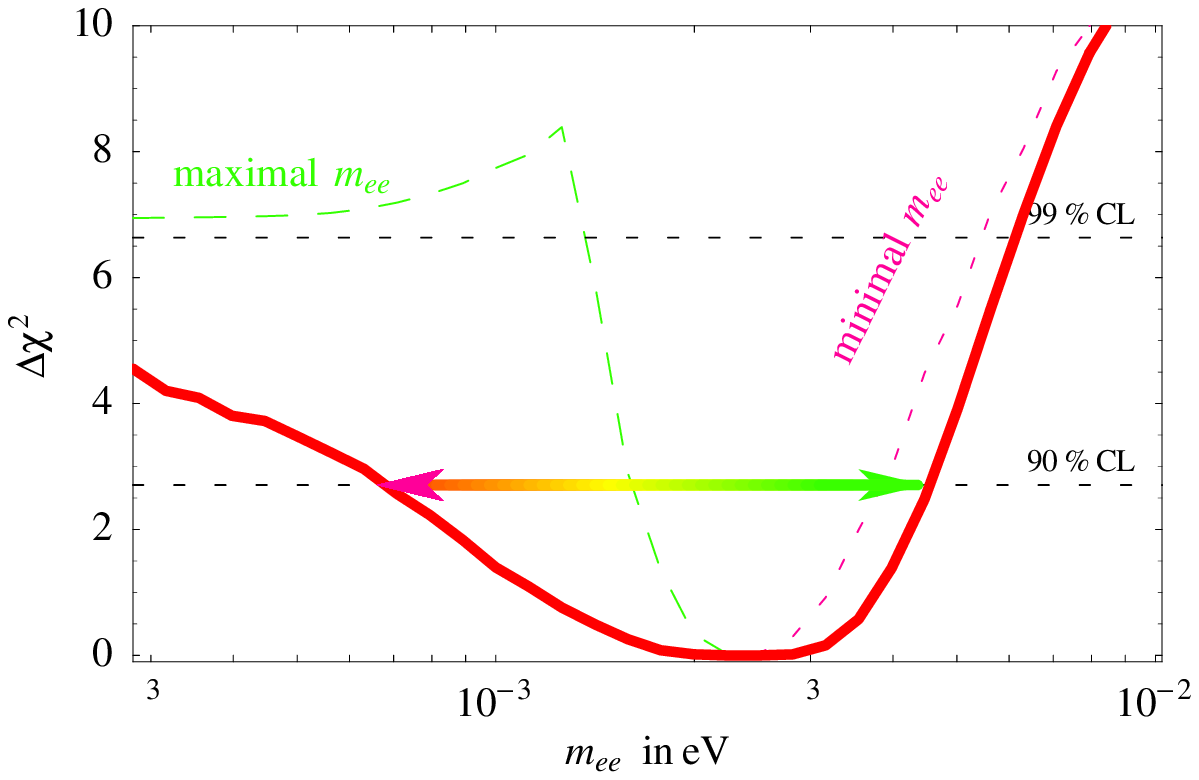}
$$
  \caption[]{\em In the case of normal hierarchy ($m_1\ll m_2\ll m_3$), we show
the present experimental
information on the solar and atmospheric
contributions to $m_{ee}$ (fig.\fig{dir}a)
and on $|m_{ee}|$ itself (fig.\fig{dir}b).
\label{fig:dir}}
\end{figure}

The crucial question is:
can this prediction be practically tested?
Non-oscillation $0\nu2\beta$ experiments can do that, {\em provided that
the `solar' and the `atmospheric' contributions to $m_{ee}$
are not comparable} \cite{mee}.
In fact, we have
$$|m_{ee}| =
|\sqrt{\Delta m^2_{12}} \sin^2 \theta_{12}\cos^2 \theta_{13} e^{2 i  \alpha}
+
\sqrt{\Delta m^2_{23}} \sin^2 \theta_{13} e^{2 i  \beta}|
\equiv
|m_{ee}^{\rm sun} + e^{2i (\beta-\alpha)} m_{ee}^{\rm atm}|$$
\noindent
CP-violating phases enter through the combination $\beta-\alpha$.
Therefore, if oscillation experiments will tell that
$m_{ee}^{\rm sun} \gg m_{ee}^{\rm atm}$,
this sub-class of see-saw models will imply that
$|m_{ee}|$ is in the narrow range
$$m_{ee}^{\rm sun} - m_{ee}^{\rm atm}
\le |m_{ee}| \le m_{ee}^{\rm sun} + m_{ee}^{\rm atm}
$$
An analogous prediction is possible in the opposite eventuality,
$m_{ee}^{\rm atm} \gg m_{ee}^{\rm sun}$:
in this case $|m_{ee}|$ would be approximately equal
to $m_{ee}^{\rm atm}$.

What present oscillation data tell us on these
eventualities?  The answer is illustrated in
fig.\fig{iso}, where we show the values of these 
two contributions in the regions of parameters 
compatible with solar and atmospheric neutrino 
oscillations. From these results
(and applying the statistical procedure
described in section~\ref{osc}) we extract the
$\chi^2$ distributions of
the solar and atmospheric contributions 
to $m_{ee}$ plotted in fig.\fig{dir}a.
We see that the largest contribution allowed by data
is the solar one:
if LMA is the solution of the solar anomaly,
solar data suggest a $m_{ee}^{\rm sun}$ in the $99\%$ CL range $1\div 10$ meV.
The {\sc Chooz} bound on $\theta_{13}$, together with the SuperKamiokande
measurement of $\Delta m^2_{23}$, give the $99\%$ CL upper bound $m_{ee}^{\rm
atm}\circa{<}3\meV$. 
Therefore the atmospheric contribution alone
is too small for giving a signal in planned experiments. 
The dominant upper bound on both parameters $m_{ee}^{\rm atm}$  and
$m_{ee}^{\rm sun}$ (rightmost lines in fig.\fig{dir}a)
is provided by {\sc Chooz}, that excludes large $\theta_{13}$ and
large $\Delta m^2_{12}$.

In fig.\fig{dir}b we show
the probability distribution of $|m_{ee}|$ (thick line),
taking into account our ignorance on the CP-violating phase $\alpha-\beta$.
The dashed lines also show the probability distribution of $|m_{ee}|$
in the special cases of maximal ($\alpha-\beta=0$) and
minimal ($\alpha-\beta=\pi/2$) interference between the
atmospheric and solar contributions.

Assuming Majorana neutrinos with normal hierarchy,
$|m_{ee}|$ is {\em well predicted and large enough}
to be measured in forthcoming $0\nu2\beta$ experiments
in two eventualities:
\begin{enumerate}
\item $\theta_{13}$ is somewhat below the {\sc Chooz} bound and
$\Delta m^2_{\rm 12}$ is
in the upper part of the LMA region;

\item $\theta_{13}$ is around the {\sc Chooz} bound, and
$\Delta m^2_{\rm 12}$ is not in
LMA region.
\end{enumerate}
In the first case it will be possible to test,
and eventually exclude, the sub-class of models that predict $m_1\ll m_2$
(the accuracy of this test will be limited by the theoretical
uncertainty on the nuclear
matrix element relevant for the $0\nu2\beta$ decay).
In the second case the actual value of
$m_1$ is irrelevant even if $m_1\sim m_2\sim(\Delta m^2_{\rm 12})^{1/2}$; however,
the case favored by the data is the first one.

\medskip

According to present data, `solar' and `atmospheric'
contributions to $m_{ee}$ can be comparable and sizable.
In this case, what we will learn from
a future very sensitive $0\nu 2\beta$ experiment?
\begin{itemize}
\item A measurement of $|m_{ee}|$ would be
equivalent to a measurement
of the Majorana CP-violating phase $\alpha-\beta$, if oscillation
parameters are known.

\item
A strong upper bound on
$|m_{ee}|$ would suggest $\alpha-\beta\approx \pi/2$ and
a correlation between oscillation parameters:
$\tan^2 \theta_{13}\approx \sin^2\theta_{12}
\sqrt{\Delta m^2_{12}/\Delta m^2_{23}}$
(or that
both $m_{ee}^{\rm sun}$
and $m_{ee}^{\rm atm}$ 
are negligibly small).

\end{itemize}
%
Concerning $\beta$-decay,
the range for $m_{\nu_e}$ is $(3\div 10)\meV$ at $90\%$ CL 
and $(2.8\div 12)\meV$ at $99\%$ CL
(a negligible $m_{\nu_e}$ can be obtained in the LOW solution with a small $\theta_{13}$).
The
largest contribution to $m_{\nu_e}$ could be the one associated with
atmospheric oscillations without contradicting present data
(see remark 3 at page~\pageref{remarks}), but would remain
too small to give any detectable effects
in planned $\beta$-decay experiments.

\begin{figure}[t]
$$\hspace{-4mm}
\includegraphics[width=16cm]{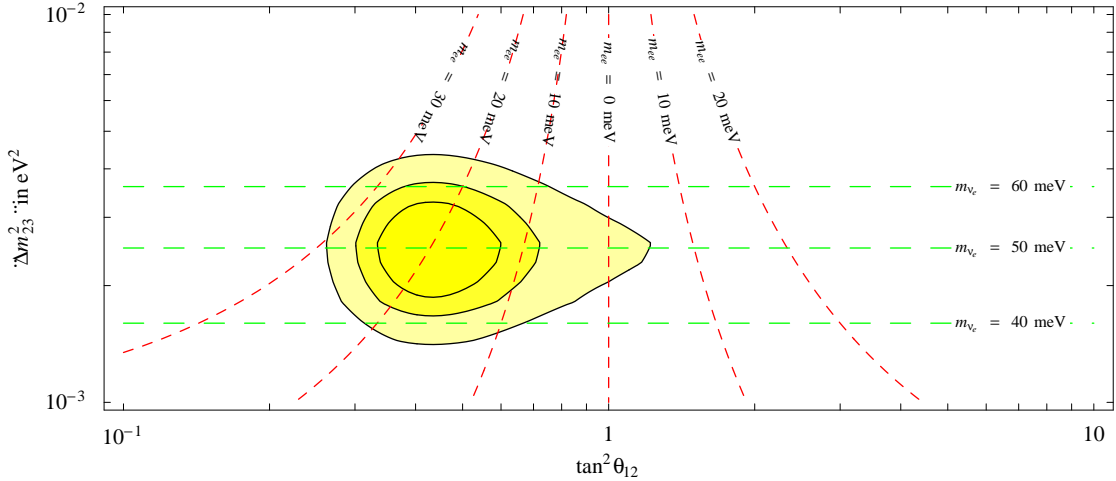}
$$
  \caption[]{\em Contour lines of the minimal (short-dashed red lines) and maximal contribution to
to $m_{ee},$ equal to $m_{\nu_e}$ (long-dashed green lines) 
in the case of inverted hierarchy.
Superimposed, the regions allowed at $68,90,99\%$ CL by oscillation data.
\label{fig:inva2}}
\end{figure}

\begin{figure}[t]
$$\hspace{-4mm}
\includegraphics[width=9cm]{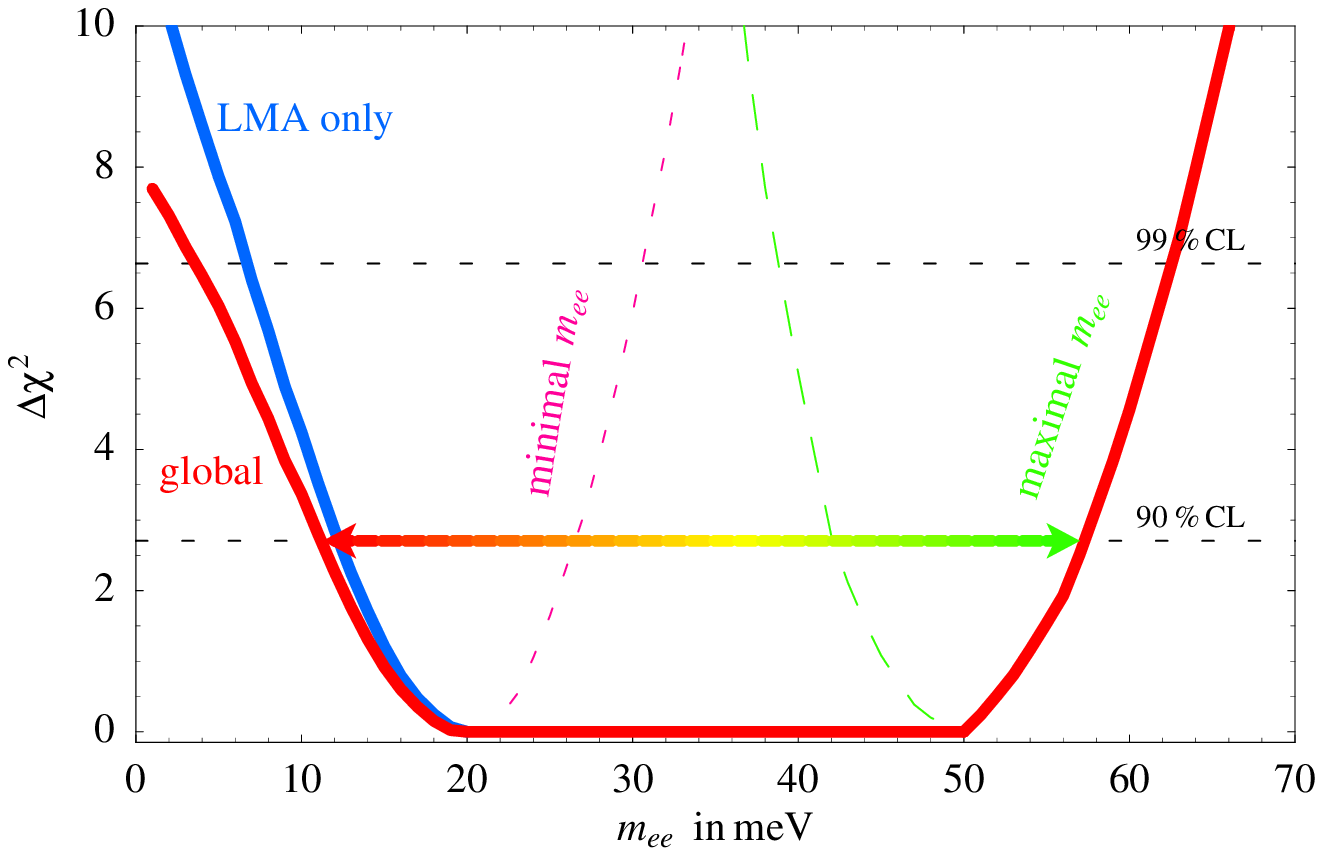}
\includegraphics[width=9cm]{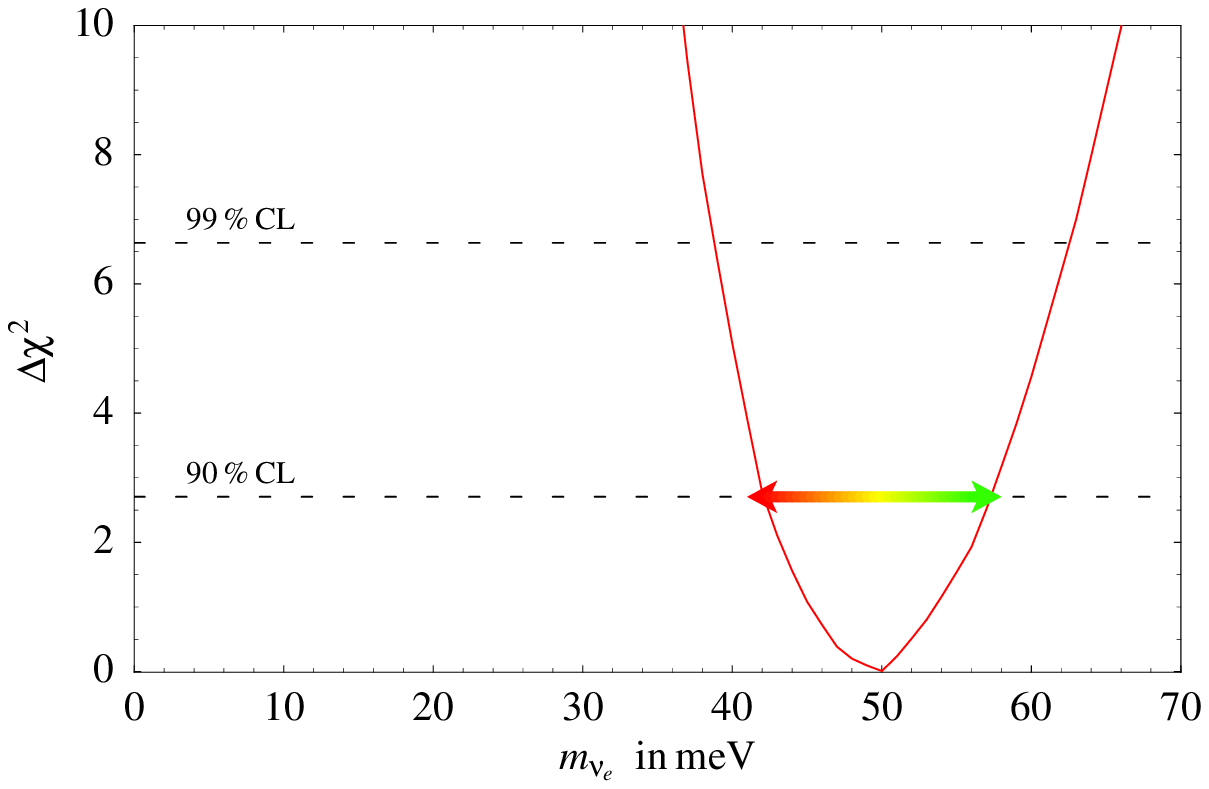}
$$
  \caption[]{\em In the case of inverted hierarchy, we plot
the probability distribution of the minimal and maximal values of
$m_{ee}$ (fig.\fig{inva}a),
and the probability distribution of $m_{\nu_e}$ (fig.\fig{inva}b).
\label{fig:inva}}
\end{figure}

\subsection{Inverted hierarchy}\label{sect:i}
In our notation `inverted hierarchy' means $m_3\ll m_1\approx m_2\approx(\Delta m^2_{32})^{1/2}$,
with $m_1$ and $m_2$ separated by the `solar' mass splitting.
The general expressions\eq{mnue} and \eq{02} simplify to
$$
m_{\nu_e}\approx 
(\Delta m^2_{23})^{1/2} \cos \theta_{13},\qquad
|m_{ee}| \approx (\Delta m^2_{23})^{1/2}
|\cos^2\theta_{12} +  e^{2i\alpha} \sin^2 \theta_{12}|\cos^2 \theta_{13}
$$
so that $m_{\nu_e}$, and the value of $m_{ee}$ maximized
with respect to Majorana phases, is
practically equal to the atmospheric mass splitting,
$m_{ee}^{\rm max} = m_{\nu_e} = (\Delta m^2_{32})^{1/2}$.
The case of inverted hierarchy seems more favourable, even from a
pessimistic point of view.
Indeed,  a non zero value
\begin{equation}\label{eq:invmin}
|m_{ee}|>
[ \Delta m^2_{32} (\cos^2 2 \theta_{12}
-\theta_{13}^4)]^{1/2} \cos\theta_{13}
\mp  \frac{\Delta m^2_{12}}{2 (\Delta m^2_{32})^{1/2}}
\sin^2\theta_{12}
\label{sure2}
\end{equation}
is now {\em guaranteed} in the two regions (corresponding to $\mp$ sign
in previous formula)
\begin{equation}
(\theta_{12}< \frac{\pi}{4} -
\frac{1}{8}\frac{\Delta m^2_{12}}{\Delta m^2_{32}}
\quad\hbox{or}\quad
\theta_{12} > \frac{\pi}{4})
\qquad\hbox{and}\qquad
\theta_{13} < \sqrt{2}
\bigg|\theta_{12} - \frac{\pi}{4}\bigg|^{1/2}
\end{equation}
up to higher orders in
$\theta_{13}$ and ${\Delta m^2_{12}}/{\Delta m^2_{32}}$.
If $\theta_{13}$ and $\Delta m^2_{12}$ can be neglected, eq.~(\ref{eq:invmin})
reduces to the well-known result
$|m_{ee}^2| > \Delta m^2_{32} \cdot \cos^2 2\theta_{12}$.
The condition on $\theta_{12}$ is satisfied
by best fit LMA oscillations.
Furthermore, as discussed in section~\ref{osc},
SN1987A data could prefer a
small $\theta_{13}\circa{<}1^\circ$
for inverted hierarchy.

Assuming a negligible $\theta_{13}$ and $\Delta m^2_{12}\ll |\Delta m^2_{23}|$, the situation is
illustrated in fig.\fig{inva2}, where we plot
contour lines of the minimal value of $m_{ee}$ (obtained for $\alpha=\pi/2$)
and of the maximal value of $m_{ee}$ (obtained for $\alpha=0$ and equal to $m_{\nu_e}$)
superimposed to the $68,90,99\%$ CL confidence regions
for the solar mixing angle and for the atmospheric mass difference.
A cancellation in $m_{ee}$ is possible for $\tan\theta_{12}\approx 1$:
such values of $\theta_{12}$ would be strongly disfavoured if
we could restrict our analysis to the LMA solution of the solar neutrino anomaly.

Applying the statistical procedure
described in section~\ref{osc}, we extract the
$\chi^2$ distributions of $m_{ee}$ and $m_{\nu_e}$.
The result is shown in fig.\fig{inva}:
oscillation data
suggest a $m_{\nu_e}$ in the $90\%$ CL range
$(40\div 57)\,\meV$
and a
$|m_{ee}|$ in the $90\%$ CL  range
$(12\div 57)\,\meV$.
However a much smaller $|m_{ee}|$ can be obtained
at slighly higher CL from patterns of solar oscillations
(mainly LOW or (Q)VO with maximal mixing)
somehow disfavoured by data, but favoured by certain
theoretical considerations.
As discussed in section~\ref{models}, inverted hierarchy
is naturally obtained from a Majorana mass matrix
of pseudo-Dirac form, that predicts
all the conditions that give a small or
zero $m_{ee}$:
$\theta_{12}\approx \pi/4$, $\theta_{13}\approx 0$ and $\alpha\approx \pi/2$.

In general, if neutrinos have Majorana masses with inverted spectrum
\begin{itemize}
\item an accurate measurement of $|m_{ee}|$ would be equivalent to a measurement
of the Majorana CP-violating phase $\alpha$, if
oscillation parameters are known;

\item a strong upper bound on $m_{ee}$ would
imply $\alpha\approx \pi/2$
and a correlation between oscillation parameters,
that probably would be practically indistinguishable
from $\theta_{12}\approx \pi/4$.

\end{itemize}

\subsection{Almost degenerate neutrinos \label{sect:d}}
In this case, the common neutrino mass is essentially
equal to the parameter $m_{\nu_e}$ probed by $\beta$-decay experiments.
This is presumably the only case that can be tested by
near future $\beta$-decay experiments,
and certainly the only case that could
affect present experiments.

We discuss in the appendix why we do not consider the recently 
claimed ``evidence for $0\nu 2\beta$''~\cite{evid} convincing. Therefore,
in this section we will limit to consider the existing {\em bound}
on $0\nu 2\beta$~\cite{HM}, and we will compute the bound on $m_{\nu_e}$ from
$0\nu 2\beta$ and oscillation data, that can be compared with the
one from $\beta$ decay. The argument goes as follows:
letting aside the region
where cancellations are possible~\cite{deg,mee},
\begin{equation}\label{eq:0deg}
\theta_{12}\neq \frac{\pi}{4}
\qquad\hbox{and}\qquad
\tan\theta_{13} < |\cos 2\theta_{12} |^{1/2}
\end{equation}
almost degenerate neutrinos lead to
a non-zero value of $m_{ee}:$
\begin{equation}
|m_{ee}| \ge m_{\nu_e} \cdot (\cos^2\theta_{13}|\cos 2\theta_{12}| -
\sin^2\theta_{13}).
\end{equation}
Thence, the question is whether we know
$\theta_{12}$ and $\theta_{13}$ sufficiently well to
exclude a cancellation.

\begin{figure}[t]
$$\includegraphics[width=100mm]{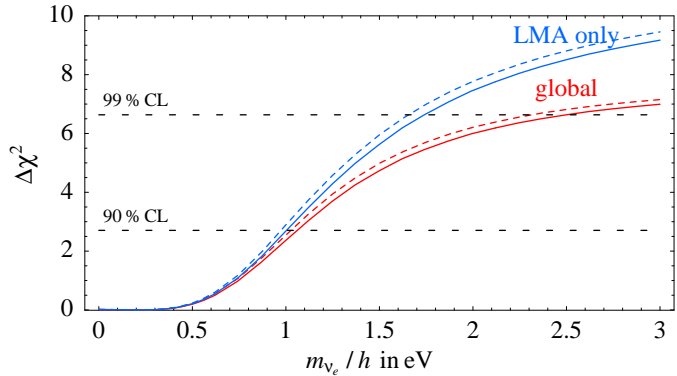} $$
\caption{\em
The present experimental bound on the mass of degenerate neutrinos
from $0\nu2\beta$ and oscillation experiments.
The dotted lines illustrate the constraint obtained setting $\theta_{13}=0.$
$h\approx 1$ parameterizes the uncertainty in the 
nuclear matrix element (see
sect.\ \protect{\ref{sect:noOscExps}}).
\label{fig:fit}}
\end{figure}

The answer is a `conditional yes', as
can be seen from fig.\fig{fit}.
There we  show the $\Delta \chi^2$
obtained combining all existing
$0\nu 2\beta$ and oscillation data,
as function of $m_{\nu_e}$.\footnote{
The dominant bound on $\theta_{13}$ comes from
the {\sc Chooz} experiment (when combined  with the
SK atmospheric constraints on  $|\Delta m^2_{\rm 23}|$).
The replacement $\theta_{13}=0$, however,
approximates reasonably  well the
accurate treatment of the {\sc Chooz} bound.}
At $90\%$ CL, this procedure yields
the interesting bound $m_{\nu_e} <1.05 h\eV$,
which is better than the limit from $\beta$-decay
if $h$ is not too large.
If we knew that LMA is the true solution of
the solar neutrino problem, we would have the 99\% CL bound
$m_{\nu_e} <1.5 h\eV$.
These considerations illustrate
the importance of obtaining a precise measurements
of the `solar' mixing angle $\theta_{12}$, and of reducing
nuclear uncertainties to get further
information on massive neutrinos.
It will be important to reconsider
the inference on neutrino mass scale $m_{\nu_e}$
from $0\nu2\beta$, as soon as we will get
more precise information on the solar mixing
angle $\theta_{12}$ from SNO, KamLAND, Borexino.
The outcome would be particularly interesting if KamLAND
will confirm the LMA solution with less-than-maximal mixing.


We summarize  the possible r\^ole of future
$\beta$-decay and $0\nu2\beta$-decay experiments as
follows
\begin{itemize}
\item If neutrinos have Majorana masses
and if future oscillation data will tell us
that eq.~(\ref{eq:0deg}) is satisfied at a high CL,
{\em $0\nu2\beta$ experiments alone will be sufficient to rule out
(or confirm)
the possibility that neutrinos have a large common mass
$m_{\nu_e}$}.

\item If neutrinos have a large common Majorana mass
$m_{\nu_e}$,
$\beta$-decay experiments will be necessary to measure it,
and $0\nu 2\beta$ experiments will permit to investigate the Majorana
phases.
Since
$V_{e3}^2$ is small,
$\alpha$ is
the only relevant phase, to a good approximation.

\end{itemize}

\section{Predictions, expectations, guesses and prejudices}\label{models}
Perhaps the simplest and more conservative theoretical explanation 
of the lightness of neutrinos is given by the idea that
neutrino masses are suppressed by the scale of lepton number
violation~\cite{generale}. This scenario is well compatible with 
unification of electroweak and strong interactions at a very high
energy scale and provides an appealing mechanism for the generation of the 
observed baryon asymmetry. Such considerations single out,
among all possible models of neutrino masses and mixings, those
characterized, at low energies, by three light Majorana neutrinos.
Given the present experimental knowledge, that favors $\Delta m^2_{23}$ as 
the leading oscillation frequency, it is also natural to define a
zeroth order approximation of the theory, where $\Delta m^2_{12}$
and $\theta_{13}$ vanish (which allows us to neglect the CP-breaking 
parameter $\phi$) whereas $\theta_{23}$ and $\theta_{12}$ are maximal.
This approximation is of course not realistic and should be regarded
only as a limiting case, possibly arising from an underlying symmetry.
Many effects can perturb this limit, such as small symmetry breaking
terms, radiative corrections, effects coming from residual
rotations needed to diagonalize the charged lepton mass matrix, or
to render canonical the leptonic kinetic terms. A detailed analysis
of these effects would require a separate discussion for each 
conceivable model, with no guarantee of being able to really
cover all theoretical possibilities. Here we will adopt a less
ambitious approach which consists in dealing only with the simplest
perturbations of the leading textures, with the hope that the results
will be sufficiently representative of the many existing models.
This point of view is only partially supported by present data.
If $\Delta m^2_{12}$ and $\theta_{13}$ were as large
as experimentally allowed, neutrino data would not clearly point
to any simple pattern  of the type considered here.

\medskip
Of course, there are many other promising
theoretical ideas on neutrino masses, 
that maintain their interest (and survive) even after 
the recent experimental developments. 
Some of these 
recent approaches insist on (certain minimal version of) grand 
unification models~\cite{babu}, and, in particular 
aim at a link with proton decay rate~\cite{pati}.
There are also attemps to find 
stricter relations with leptogenesis~\cite{anjan};
or with low energy lepton flavor violating
processes in supersymmetric models~\cite{aleEbarb}; {\em etc}.
We do not want to underestimate 
the validity of these ideas,
and we instead hope that some or all of them
will find an important role in a future
theory of massive neutrinos. Rather, it should be clear that 
our approach to the theory of neutrino masses is guided by simple and
phenomenological considerations, to the aims of exploiting
regularities in the observable parameters of 
massive neutrinos, and of making a guess 
about $m_{ee}$ and $\theta_{13}$.

\subsection{Normal hierarchy}\label{sect:nth}
We begin with the hierarchical neutrino spectrum, whose leading
texture, up to redefinitions of the field phases, 
in the mass-eigenstate bases of charged leptons,
is given by
\be
m_\nu
=m
\left(
\begin{array}{ccc}
0& 0& 0\\
0& 1& 1\\
0& 1& 1
\end{array}
\right)~~~.
\label{hier}
\ee
This structure might be due to a peculiar 
structure of the mass matrix of the right-handed neutrinos, 
of the (Dirac) Yukawa couplings of the neutrinos, 
or even of the charged leptons
in a certain flavor basis \cite{proposals}; this 
may arise, for instance, within a U(1) flavor symmetry \cite{u1}
under which $\nu_\mu$ and $\nu_\tau$ are neutral and 
$\nu_e$ possesses a non-vanishing charge~\cite{nhu1,af1,af2}. 
The determinant of the block $(2,3)$ can be forced to vanish
(so that one naturally obtains $\Delta m^2_{12}\ll\Delta m^2_{23}$)
by making use of the see-saw mechanism,
if one of the heavy right-handed neutrinos
has larger Yukawa couplings and/or is lighter than the other ones~\cite{king}.\footnote{The 
suggestion to
interpret  the data on massive neutrinos by assuming a number of 
right handed neutrinos as small as possible is quite old, 
see~\cite{surprise}. 
Note however that in this work it
was missed  the fact that the model 
with 2 right-handed neutrinos is predictive,
as we discuss in the following.}
The resulting neutrino mass matrix has approximately rank 
one. This setup can be implemented
within the framework of a U(1) symmetry~\cite{af1,af2}, 
which, however, only predicts 
a generically large, not necessarily maximal $\theta_{23}$. 
Another possibility to generate the leading texture in eq.~(\ref{hier})
is at 1 loop~\cite{rparl} or at tree level~\cite{rpart} 
in supersymmetric models that violate both lepton number and R-parity. 
An important feature of (\ref{hier}) is that $\theta_{12}$ is undetermined
at leading order. 

To get a realistic mass matrix, we consider 
deviations from the symmetric limit in (\ref{hier}), parametrized by:
\be
m_\nu
=m
\left(
\begin{array}{ccc}
\delta& \epsilon& \epsilon\\
\epsilon& 1+\eta& 1+\eta\\
\epsilon& 1+\eta& 1+\eta
\end{array}
\right)
\label{hier1}~~~,
\ee
where $\delta$, $\epsilon$ and $\eta$ denote small ($\ll 1$)
real parameters, defined up to coefficients of order one that can
differ in the various matrix elements. 
The mass matrix in (\ref{hier1}) does not describe the most
general perturbation of the zeroth order texture  (\ref{hier}).
We have implicitly assumed a symmetry between $\nu_\mu$ 
and $\nu_\tau$ which is preserved by the perturbations,
at least at the level of the order of magnitudes.
It is difficult to understand the
precise origin of these small deviations.
However, it is possible to construct models based
on a spontaneously broken U(1) flavor symmetry, where  
$\delta$, $\epsilon$ and $\eta$ are given by positive powers
of one or more symmetry breaking parameters. Moreover, by playing
with the U(1) charges, we can adjust, to certain extent, 
the relative hierarchy between $\eta$, $\epsilon$ and 
$\delta$~\cite{isa}.  

Another example is given by those models where
the neutrino mass matrix elements are dominated,
via the see-saw mechanism, by the exchange of two right-handed
neutrinos~\cite{king2}. Since the exchange of a single right-handed neutrino
gives a successful zeroth order texture, we are encouraged to
continue along this line.
Thus, we add a sub-dominant contribution of a second
right-handed neutrino,
assuming that the third one
gives a negligible contribution
to the neutrino mass matrix, because
it has much smaller Yukawa couplings
or is much heavier than the first two.
The Lagrangian
that describes this plausible sub-set of see-saw models,
written in the mass eigenstate
basis of right-handed neutrinos and charged leptons,
is
$$\Lag = \lambda_\ell L_\ell Nh + \lambda'_\ell L_\ell N'h +
\frac{M}{2} N^2 + \frac{M'}{2} N^{\prime 2}\qquad\Rightarrow\qquad
m_{\ell\ell'} \propto
\frac{\lambda_\ell \lambda_{\ell'}}{M} +
\frac{ \lambda'_\ell \lambda'_{\ell'}}{M'}$$
where $\ell,\ell'=\{e,\mu,\tau\}$.
The pattern of perturbations in eq. (\ref{hier1}) can be reproduced
if $\lambda_e\ll \lambda_\mu\approx\lambda_\tau$ and $\lambda'_\mu\approx
\lambda'_\tau$. Even though the number of see-saw parameters is larger 
than the number of low-energy observables, there is one neat {\em prediction}:
$$\det m = 0\qquad \hbox{i.e.}\qquad m_1=0~~~.$$
(We recall that the possibilities to test this case were discussed
at length in section~\ref{sect:n}).

Let us come back to the mass matrix $m_\nu$ of eq.\ (\ref{hier1}).
After a first rotation by an angle $\theta_{23}$ close to 
$\pi/4$ and a second rotation with $\theta_{13}\approx \epsilon$, we get
\be
m_\nu
\approx m
\left(
\begin{array}{ccc}
\delta+\epsilon^2& \epsilon&0 \\
\epsilon& \eta& 0\\
0& 0& 2
\end{array}
\right)
\label{hier2}~~~,
\ee 
up to order one coefficients in the small entries.
To obtain a large solar mixing angle, we need $|\eta-\delta|\circa{<}
\epsilon$. In realistic models there is no reason for a cancellation 
between independent perturbations and thus we assume 
$\delta\circa{<}\epsilon$ and $\eta\circa{<}\epsilon$, separately.  

Consider first the case $\delta\approx \epsilon$ and $\eta\circa{<}\epsilon$.
The solar mixing angle $\theta_{12}$ is large
but not maximal, as preferred by the large angle MSW solution.
We also have $\Delta m^2_{23}\approx 4 m^2$, $\Delta m^2_{12}
\approx m^2 \epsilon^2$ and 
\be
m_{ee}\approx \sqrt{\Delta m^2_{12}}~~~,
\label{bestnh}
\ee
the largest possible prediction, in the case of normal hierarchy,
in agreement with the results in section~\ref{sect:n}.

If $\eta\approx \epsilon$ and $\delta\ll\epsilon$,  
we still have a large solar mixing angle and
$\epsilon^2\approx\Delta m^2_{12}/\Delta m^2_{23}$, as before.
However $m_{ee}$ will be much smaller than the estimate in (\ref{bestnh}).
Unfortunately, this is the case of the models based on the
above mentioned U(1) flavor symmetry that, at least in its
simplest realization, tends to predict $\delta\approx \epsilon^2$.
In this class of models we find 
\be
m_{ee}\approx \sqrt{\Delta m^2_{12}}
\left(
\frac{\Delta m^2_{12}}{\Delta m^2_{23}}
\right)^{\frac{1}{2}}~~~, 
\label{nh2}
\ee
below the sensitivity of the next generation of planned 
experiments. It is worth to mention that in both cases
discussed above, we have 
\be
\theta_{13}\approx 
\left(\frac{\Delta m^2_{12}}{\Delta m^2_{23}}\right)^{\frac{1}{2}}~~~,
\label{ue3nh}
\ee
which might be very close to the present experimental 
limit.

If both $\delta$ and $\eta$ are much smaller than
$\epsilon$, the $(1,2)$ block of $m_\nu$ has an approximate pseudo-Dirac
structure and the angle $\theta_{12}$ becomes maximal.
This situation is typical of models where leptons have U(1) charges
of both signs whereas the order parameters of U(1) breaking
have all charges of the same sign~\cite{af2}. 
We have two eigenvalues approximately given by $\pm m~\epsilon$.
As an example, we consider the case where $\eta=0$ and 
$\delta\approx\epsilon^2$. We find $\sin^2 2\theta_{12}\approx
1-\epsilon^2/4$, $\Delta m^2_{12}\approx m^2 \epsilon^3$
and
\be
m_{ee}\approx \sqrt{\Delta m^2_{12}}
\left(
\frac{\Delta m^2_{12}}{\Delta m^2_{23}}
\right)^{\frac{1}{6}}~~~. 
\label{nh3}
\ee
In order to recover the large angle MSW solution we would
need a relatively large value of $\epsilon$. This is in general
not acceptable because, on the one hand the presence of
a large perturbation raises doubts about the consistency 
of the whole approach and, on the other hand, in existing models
where all fermion sectors are related to each other, $\epsilon$
is never larger than the Cabibbo angle. We are then forced to
embed the case under discussion within the LOW solution,
where the solar frequency is much smaller and, as a consequence,
$m_{ee}$ is beyond the reach of the next generation of experiments.

\subsection{Inverted hierarchy}
If the neutrino spectrum has an inverted hierarchy, the leading
texture depends on the relative phase $2 \alpha$ between the two non-vanishing 
eigenvalues. 
In this case theory can give significant restrictions,
since one has to explain why two neutrinos are degenerate.
This degeneracy can be more easily explained
for special values of the relative phases.
In particular,
when $\alpha=\pi/2$, the resulting texture has an exact
symmetry under the combination $L_e-L_\mu-L_\tau$~\cite{bhsw} 
and reads:
\be
m_\nu
=m
\left(
\begin{array}{ccc}
0& 1& 1\\
1& 0& 0\\
1& 0& 0
\end{array}
\right)~~~.
\label{invh}
\ee
The Abelian symmetry $L_e-L_\mu-L_\tau$ would allow different
(1,2) and (1,3) entries, that are chosen equal in (\ref{invh})
to recover a maximal $\theta_{23}$. Independently from
the relative size of these entries, we have 
$\theta_{12}=\pi/4$, $\theta_{13}=0$, $\Delta m^2_{23} = 2 m^2$,
$\Delta m^2_{12}=0$. 
Notice that an exact $L_e-L_\mu-L_\tau$ symmetry implies $m_{ee}=0$.
If it were possible to
find a symmetry realizing the inverted hierarchy 
with $\alpha\ne\pi/2$, then we could avoid $m_{ee}=0$.
We add small perturbations according to:
\be
m_\nu
=m
\left(
\begin{array}{ccc}
\delta& 1& 1\\
1& \eta& \eta\\
1& \eta& \eta
\end{array}
\right)~~~.
\label{invh1}
\ee
The perturbations leave $\Delta m^2_{23}$ and $\theta_{23}$ unchanged,
in first approximation. We obtain $\theta_{13}\approx\eta$,
$\tan^2\theta_{12}\approx 1+\delta+\eta$ and 
$\Delta m^2_{12}/\Delta m^2_{23}\approx \eta+\delta$, where
coefficients of order one have been neglected.

There is a well-known difficulty of this scenario to reproduce
the large angle MSW solution~\cite{bhsw,bd}. 
Indeed, barring cancellation between
the perturbations, in order to obtain a $\Delta m^2_{12}$ close to the best fit LMA value, 
$\eta$ and $\delta$ should be smaller
than about 0.1 and this keeps the value of $\sin^2 2\theta_{12}$ very close  to 1, 
somewhat in disagreement with global fits of solar data~\cite{sunfit}. 
Even by allowing for a $\Delta m^2_{12}$ in the upper range of the LMA solution,
or some fine-tuning between $\eta$ and $\delta$, we would need
large values of the perturbations to fit the LMA solution. 
On the contrary, the LOW solution can be accommodated, but, 
even in the optimistic case $\delta\gg\eta$, we obtain:
\be
m_{ee}=\frac{1}{2}\sqrt{\Delta m^2_{12}} 
\left(\frac{\Delta m^2_{12}}{\Delta m^2_{23}}\right)^{\frac{1}{2}}~~~,
\label{ue3ih1}
\ee
too small to be detected by planned experiments. This is an example 
where the underlying symmetry forces a cancellation between potentially
large contributions to $m_{ee}$, which persists also 
after the inclusion of
the perturbations. In fact the combination $m_1 \cos^2\theta_{12}+
m_2 e^{2 i\alpha}\cos^2\theta_{12}$ is not of order $m$, as we could
naively expect, but of order $m~ \delta$. The largest allowed value
for $\theta_{13}$ is
\be
\theta_{13}\approx 
\frac{\Delta m^2_{12}}{\Delta m^2_{23}}~~~,
\label{ue3ih2}
\ee
which, for the LOW solution, is practically unobservable.


\subsection{Almost degenerate neutrinos}\label{th:d}
Also in the case of degenerate spectrum, the zeroth order texture
depends on the relative phases between the eigenvalues and this
dependence leads to widely different expectation for $m_{ee}$. 
We have two limiting cases: one for $\alpha=0$, and another one with
$\alpha=\pi/2$. In the first case 
$m_{ee}$ will be comparable to the average neutrino mass, while in the
second case the large mixing in the solar sector tends to
deplete the $m_{ee}$ entry \cite{fg}. There are examples of both these
possibilities among the theoretically motivated models and here we
will discuss two representative scenarios. It should be said 
that it is more difficult to accommodate a degenerate neutrino spectrum 
in a model of fermion masses, than a spectrum with normal or inverted 
hierarchy.
The neutrino degeneracy should be made compatible with the observed 
hierarchy in the charged fermion sectors, and this is not an easy
task. Once the degeneracy is achieved, at leading order, it
should be preserved by renormalization group evolution
from the energy scale where neutrino masses originate down
to the low energy scales where observations take place.
The stability under these corrections may impose severe restrictions
on the theory, especially if a very small mass splitting is
required by the solar neutrino oscillations \cite{stab,radiative}. 
There is also a 
generic difficulty in embedding
such a model in a grand unified context where the particle
content often includes right-handed neutrinos. Indeed, it is
difficult to see how a presumably hierarchical Dirac mass matrix
combines with a Majorana mass matrix for the right-handed
neutrinos to give rise, via the see-saw mechanism, to a degenerate
neutrino spectrum. One idea that partially overcomes some
of the above-mentioned problems is that of flavor democracy \cite{fd},
where an approximate $S_{3L}\otimes S_{3R}$ symmetry 
leads to the following pattern of charged fermion mass matrices:
\be
m_{f}
={\hat m}_{f}
\left(
\begin{array}{ccc}
1& 1& 1\\
1& 1& 1\\
1& 1& 1
\end{array}
\right)+\delta m_{f}~~~,\qquad f=\{\hbox{e,u,d}\}
\label{deg1}
\ee
where $\delta m_{f}$ denotes small symmetry breaking terms.
The corresponding spectrum is hierarchical and the smallness
of quark mixing angles can be explained by an almost complete
cancellation between the two unitary left transformations 
needed to diagonalize $m_{\rm u}$ and $m_{\rm d}$. The neutrino mass matrix has
the general structure:
\be
m_{\nu}
=m
\left[
\left(
\begin{array}{ccc}
1& 0& 0\\
0& 1& 0\\
0& 0& 1
\end{array}
\right)
+ r
\left(
\begin{array}{ccc}
1& 1& 1\\
1& 1& 1\\
1& 1& 1
\end{array}
\right)
\right]
+\delta m_{\nu}~~~,
\label{deg2}
\ee
where $r$ is an arbitrary parameter and $\delta m_{\nu}$ represents
a small correction to the symmetric limit. The freedom associated to
$r$, $\delta m_{\nu}$ and $\delta m_{\rm e}$ allows us to select an almost 
degenerate and diagonal $m_\nu$ and to extract the large atmospheric
and solar mixing angles from the charged lepton sector. 
However, a non-vanishing $\Delta m^2_{23}$ and maximal $\theta_{12}$,
$\theta_{23}$ angles are not determined by the symmetric limit,
but only by a specific choice of the parameter $r$ and of 
the perturbations,
that cannot be easily justified on theoretical grounds.
If, for example, we choose $\delta m_{\nu}={\rm diag}(0,\epsilon,\eta)$
with $\epsilon<\eta\ll 1$ and $r\ll \epsilon$, the solar and
the atmospheric oscillation frequencies are determined by
$\epsilon$ and $\eta$, respectively. The mixing angles are entirely
due to the charged lepton sector. A diagonal $\delta m_{\rm e}$
will give rise to an almost maximal $\theta_{12}$, $\tan^2\theta_{23}
\approx 1/2$ and $\theta_{13}\approx \sqrt{m_e/m_\mu}$.
By going to the basis where the charged leptons are diagonal,
we can see that $m_{ee}$ is close to $m$ and independent from
the parameters that characterize the oscillation phenomena.
Indeed $m$ is only limited by the
$0\nu2\beta$ decay. The parameter
$r$ receives radiative corrections~\cite{fdrad} that, at leading
order, are logarithmic and proportional to the square of the $\tau$ 
lepton Yukawa coupling. It is important to guarantee that this correction 
does not spoil the relation $r\ll \epsilon$, whose violation would lead 
to a completely different mixing pattern. This raises a `naturalness'
problem for the LOW solution. 
It would be desirable to provide a more sound basis for the choice of
small perturbations in this scenario that is quite favorable to
signals both in the $0\nu2\beta$ decay and in sub-leading
oscillations controlled by $\theta_{13}$.

If $\alpha=\pi/2$, we could have $m_{ee}$ well below the average
neutrino mass or even below the solar oscillation frequency.
This is exemplified by the following leading texture
\be
m_\nu
=m
\left(
\begin{array}{ccc}
0& \frac{1}{\sqrt{2}}& \frac{1}{\sqrt{2}}\\
\frac{1}{\sqrt{2}}& \frac{1}{2}(1+\eta)& -\frac{1}{2}(1+\eta)\\
\frac{1}{\sqrt{2}}& -\frac{1}{2}(1+\eta)& \frac{1}{2}(1+\eta)
\end{array}
\right)~~~,
\label{deg3}
\ee
where $\eta\ll 1$, corresponding to an exact bimaximal mixing, with eigenvalues
$m_1=m$, $m_2=-m$ and $m_3=(1+\eta) m$. This texture has been
proposed in the context of a spontaneously broken SO(3) flavor symmetry
and it has been studied to analyze the stability of the degenerate
spectrum against radiative corrections~\cite{radiative}. 
We add small perturbations
to (\ref{deg3}) in the form:
\be
m_\nu
=m
\left(
\begin{array}{ccc}
\delta& \frac{1}{\sqrt{2}}& \frac{1}{\sqrt{2}}(1-\epsilon)\\
\frac{1}{\sqrt{2}}& \frac{1}{2}(1+\eta)& -\frac{1}{2}(1+\eta-\epsilon)\\
\frac{1}{\sqrt{2}}(1-\epsilon)& -\frac{1}{2}(1+\eta-\epsilon)& 
\frac{1}{2}(1+\eta-2\epsilon)
\end{array}
\right)~~~,
\label{deg4}
\ee
where $\epsilon$ parametrizes the leading flavor-dependent 
radiative corrections (mainly induced by the $\tau$ Yukawa coupling)
and $\delta$ controls $m_{ee}$. We first discuss the case
$\delta\ll \epsilon$. To first approximation $\theta_{12}$ remains maximal, 
disfavouring an interpretation in the framework of the LMA
solution. We get $\Delta m^2_{12}\approx m^2 \epsilon^2/\eta$ 
and
\be
\theta_{13}\approx 
\left(\frac{\Delta m^2_{12}}{\Delta m^2_{23}}
\right)^{{1}/{2}}~~~,~~~~~~~~~~~~~
m_{ee}\ll m~\left(\frac{\Delta m^2_{23}~\Delta m^2_{12}}{m^4}\right)^
{1/2}~~~.
\label{deg5}
\ee
If we assume $\delta\gg \epsilon$, we find $\Delta m^2_{12}\approx 
2 m^2 \delta$, $\theta_{23}=\pi/4$, $\sin^2 2\theta_{12}=1-\delta^2/8$
and $\theta_{13}=0$.
Also in this case the solar mixing angle is
too close to $\pi/4$ to fit the large angle MSW solution.
We get:
\be
m_{ee}\approx \frac{\Delta m^2_{12}}{2 m}~~~,
\label{deg6}
\ee
too small for detection if the average neutrino mass $m$ is
around the eV scale.
This rather extreme example shows that there is no guarantee for
$m_{ee}$ to be close to the range of experimental interest, 
even with degenerate neutrinos where the involved masses 
are much larger than the oscillation frequencies.   

\medskip

We conclude this section with a comment on a relevant issue 
raised in~\cite{EvidTh}: would a measurement $m_{ee}\sim \eV$ 
imply the LMA solution, since in such a case the other LOW, (Q)VO solutions 
are unstable under radiative corrections?
The answer is no. Ref.~\cite{radiative} showed examples
of degenerate neutrino spectra with $m_{ee}=0$, $\theta_{13}=0$ and $\theta_{12}=\pi/4$
where $\Delta m^2_{12}$ is radiatively 
generated only at order $\lambda_\tau^4$,
rather than at order $\lambda_\tau^2$ as na\"\i{}vely expected.
It is easy to find examples with $m_{ee}\neq 0$.
The basic observation is that the results in~\cite{radiative} continue to hold 
if the neutrino mass matrix has the from in eq.~(5) of~\cite{radiative}
in a flavour basis where the charged lepton mass matrix
is not fully diagonal because has a non vanishing 12 entry.
The flavour rotation in the 
12 plane that leads to the usual basis of $e,\mu,\tau$ mass eigenstates,
generates non-vanishing $m_{ee}$ and $\theta_{13}$
and shifts $\theta_{12}$ from $\pi/4$.
Proceeding along the lines of~\cite{radiative} one concludes that in
models where
$$\alpha=\frac{\pi}{2},\qquad
\sin \theta_{13} =\tan\theta_{23}\tan(\theta_{12}-\pi/4)\qquad\hbox{so that}\quad
|m_{ee}| \simeq \frac{2 \theta_{13}}{\tan\theta_{23}}\, m+{\cal O}(\theta_{13}^2 )$$ 
radiative corrections generate a solar mass splitting
compatible with the LOW and (Q)VO solutions of the solar anomaly; 
at the same time,
$m_{ee}$ can be relatively large without 
violating the {\sc Chooz} bound.

\section{Conclusions}\label{conclusions}
As recalled in section~\ref{osc},
the experimental program on neutrino oscillations is well
under way and looks promising.
We reviewed how it should
be possible to access and reliably measure at least 4 of the 6
oscillation parameters in the near future.
KamLand and Borexino should finally identify the true solution
of the solar anomaly.
MiniBoone will test the LSND anomaly.
K2K, Minos and CNGS should reduce the error on the atmospheric parameters,
and probe values of $\theta_{13}$ below its present bound.
Further experimental progresses and improvements 
(maybe culminating in a neutrino factory) seem 
feasible in longer terms, after the first generation of 
long-baseline experiments, 
although we cannot exclude that effects due to $\theta_{13}$
or CP violation are too small to ever be observed.

The fact that oscillations are becoming an established fact
can only reinforce
the motivation for other approaches to
massive neutrinos, and primarily $\beta$ and
neutrino-less double $\beta$ ($0\nu 2 \beta$) decay experiments,
as discussed in section \ref{sect:noOscExps}.
Furthermore, $0\nu2\beta$ decay seems to be the only 
realistic possibility to
investigate the Majorana nature of neutrinos. In this
connection it is exciting that there are proposals to reach the
sensitivity to the mass scales suggested by oscillations,
$|m_{ee}^2|\sim \Delta m^2_{ij}$, but at the same time, it is sad
that we cannot exclude that $m_{ee}$ is negligibly small.
In fancier terms, it could be that
neutrinos have a `hidden Majorana nature'; namely, their mass could
be of Majorana type,
but it might be very tough to prove that
property experimentally.

\medskip

\begin{figure}
$$\includegraphics[width=10cm,height=6cm]{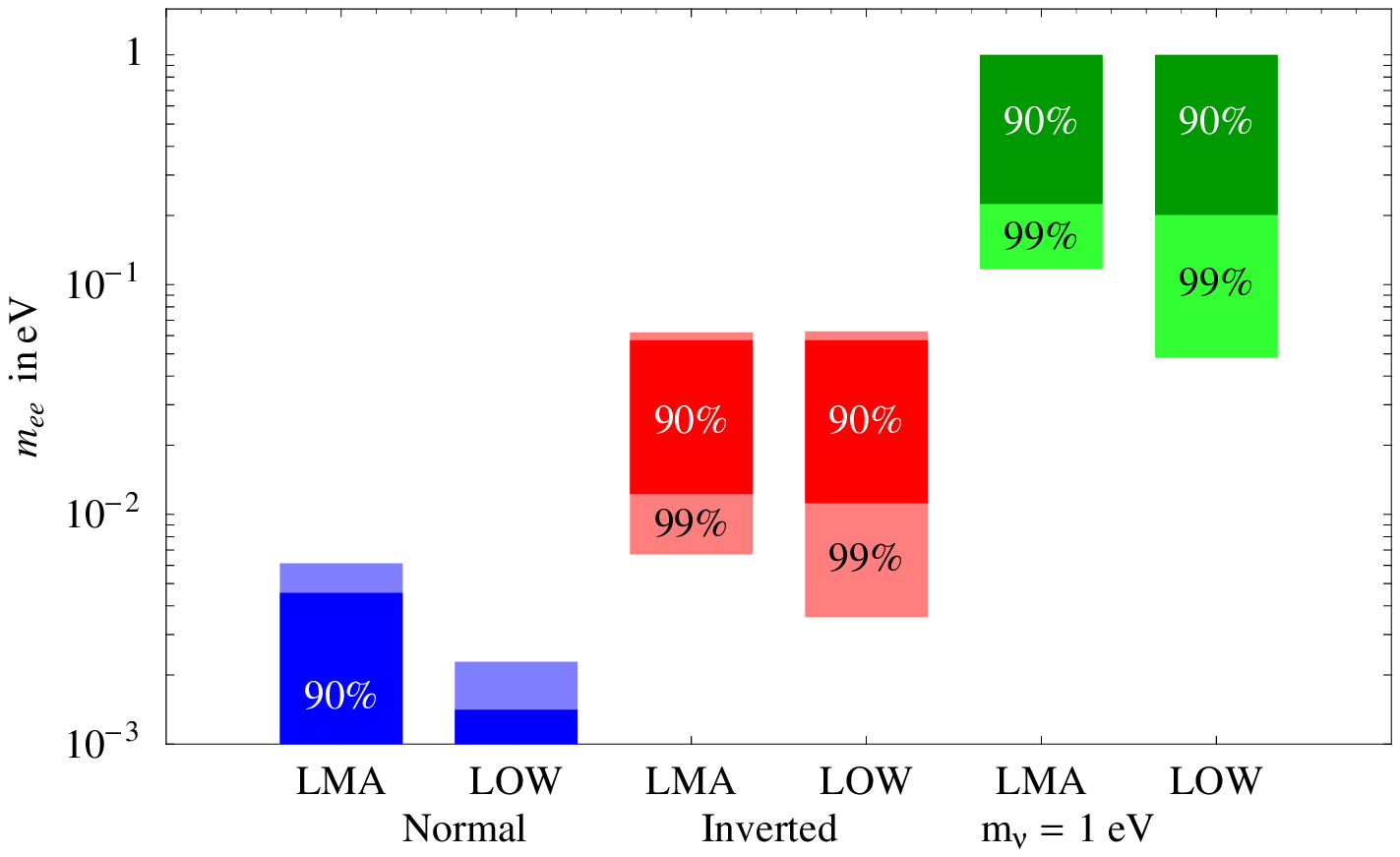}\hspace{1cm}
\includegraphics[width=6.1cm,height=6.1cm]{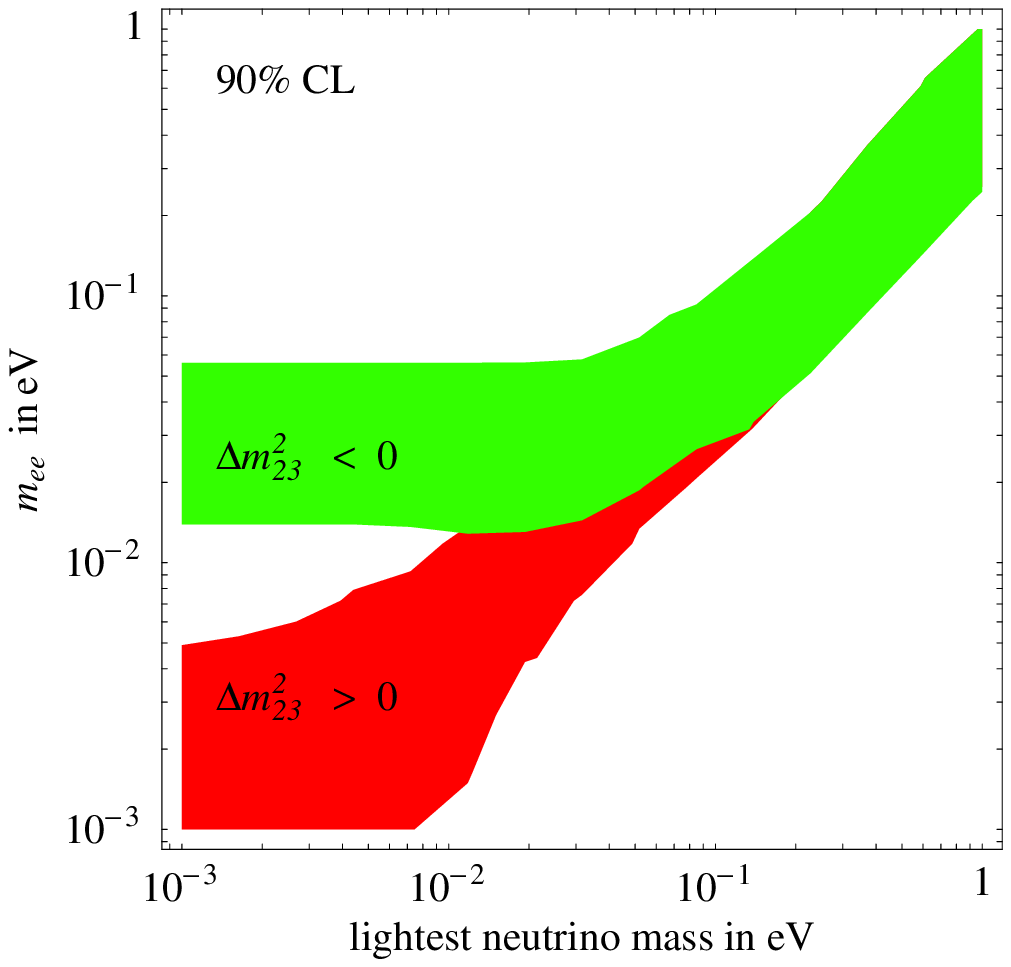}$$
\caption{\em In fig.\fig{summary}a we show the {\bf 90\% and 99\% CL ranges of $\mb{m}_{ee}$} in the
cases of  normal hierarchy (i.e.\ $0=m_1\ll m_2\ll m_2$, so that $\Delta m^2_{23}>0$)
inverted hierarchy (i.e.\ $0=m_3\ll m_1\approx m_2$, so that $\Delta m^2_{23}<0$)
and almost degenerate neutrinos at $1\eV$ (a value chosen for illustration).
LOW collectively denotes large mixing angle solutions other than LMA.
In fig.\fig{summary}b,
without distinguishing LMA or LOW, we plot the $90\%$ CL range for $m_{ee}$
as function of the lightest neutrino mass,
thereby covering all spectra.
\label{fig:summary}}
\end{figure}

We have computed the precise ranges
of $|m_{ee}|$ in the cases of 
normal and inverted hierarchy.
\begin{itemize}
\item In the case of {\bf normal hierarchy} (i.e.\ $m_1\ll m_2\ll m_3$, see
section~\ref{sect:d}) we get 
the $90\%$ CL range $|m_{ee}|=(0.7\div 4.6) \meV$.
If $\theta_{13}$ is somewhat below its present bound 
and LMA is the solution of the solar neutrino
anomaly,  this case gives a testable prediction:
$m_{ee} \approx (\Delta m^2_{12})^{1/2}\sin^2 \theta_{12}$.

\item In the case of {\bf inverted hierarchy} (i.e.\ $m_3\ll m_1\approx m_2$, see
section~\ref{sect:i}) we get the $90\%$ CL range
$|m_{ee}|=(12\div 57) \meV$.
However, a much smaller $m_{ee}$ is allowed at somewhat higher CL
and suggested by theoretical considerations.
\end{itemize}
These results are summarized in fig.\fig{summary}a, where we also show the 
ranges of $m_{ee}$,
restricted to the LMA and to the other large mixing angle  
solutions of the solar anomaly (collectively denoted as `LOW').
The SMA solution is only allowed at  higher CL than the ones we consider.

Beyond these two special cases (where the lightest neutrino 
has a little or negligible mass),
there is a family of more generic spectra
(sometimes named `partial hierarchy', `partial degeneracy', \ldots)
conveniently parametrised by the lightest neutrino mass, $m_{\rm min}$.
In fig.\fig{summary}b we show how the $90\%$ CL range for $m_{ee}$
varies as a function of $m_{\rm min}$
(note that this figure improves on traditional  
plots in $m_{\rm min}-m_{ee}$ plane \cite{mee}, which are 
done at {\em fixed} values of the oscillation parameters).
When $m_{\rm min}$ is larger than the oscillation scales
(so that neutrinos are almost degenerate at a common mass $m_{\nu_e}$)
we get the $90\%$ range $m_{ee} = (0.17\div 1)\ m_{\nu_e}$
(see fig.\fig{summary}a for more detailed results).
We also considered 
the implications of present data:
\begin{itemize}
\item In the case of {\bf almost degenerate} 
neutrinos (section~\ref{sect:d})
we converted the present  $0\nu2\beta$ bound
($m_{ee}<0.38h\eV$ at 95\% CL; $h\sim 1$ parameterizes
nuclear uncertainties, see section \ref{sect:noOscExps})
into a $90\%$ CL bound on their common mass, $m_{\nu_e} < 1.05h\eV$, competitive
with the direct $\beta$-decay bound.

\end{itemize}
This bound could become stronger with future oscillation data.
In particular, if KamLAND will confirm
the LMA solution of the solar neutrino anomaly,
it will be important to update the inferences on $m_{ee}$
performed in the present work.
Within the three Majorana neutrino
context we assumed, existing and
planned $\beta$ decay experiments can
see a signal only if neutrinos are almost degenerate.
However,
future $0\nu2\beta$ experiments alone could rule out this possibility,
if future oscillation data will safely tell us
that $|\cos2\theta_{12}|>\tan^2\theta_{13}$.
If neutrinos indeed have a large common Majorana mass,
only with both
$\beta$ and $0\nu2\beta$ experiments will it be possible
to measure it, and learn on relevant Majorana phase.

A recent paper~\cite{evid} 
claims a $0\nu2\beta$ evidence for almost degenerate neutrinos.
In the appendix we reanalysed the data on which such claim is based
and proposed a statistically fair way of extracting the 
signal from the background. Assuming that the relevant 
sources of background are the ones discussed in~\cite{evid},
we get a $1.5\sigma$ hint (or less, depending on the data-set used)
for $0\nu2\beta$ from the published data.
In section~\ref{th:d} we provide a counterexample to the statement that 
measuring $m_{ee}\sim \eV$ would
imply the LMA solution:
the other LOW, (Q)VO solutions 
would not necessarily be unstable under radiative corrections.

\medskip

All $m_{ee}$ ranges obtained above 
have been derived allowing for 
cancellations between the different contributions to $m_{ee}$.
One could be tempted to obtain more optimistic results by
invoking `naturalness' arguments to
disfavour large cancellations.
As discussed in section~\ref{How?}, 
since $m_{ee}$ is a theoretically clean quantity
(the $ee$ element of the neutrino mass matrix)
some flavour symmetry could force it to be small,
giving rise to {\em apparently} unnatural cancellations
when $m_{ee}$ is written in terms of the 
phenomenological parameters.
Phenomenology alone implies only
the relatively weak results summarized above.
%

For this reason in section~\ref{models} we have 
estimated $m_{ee}$ and $\theta_{13}$
in models of neutrino masses. We considered few
theoretically well-motivated textures, that
existing models can reproduce in some appropriate limit, and we
added small perturbations to mimic the realistic 
case. Even though this procedure cannot of course cover all possibilities,
it illustrates, for each type of spectrum, the typical theoretical
expectations for $m_{ee}$ and $\theta_{13}$ and provides
examples where flavour symmetries force a small $m_{ee}$. 
Qualitatively we can summarize our analysis by saying
that values of $m_{ee}$ above 10 meV are certainly
possible for models of quasi-degenerate neutrinos, 
but they would come as unexpected in
models characterized  by normal or inverted hierarchy.
In models realizing the LMA solutions, independently 
from the type of spectrum, the angle $\theta_{13}$ is usually
close to its present experimental limit.

\paragraph{Acknowledgment}
We are very grateful to
G.~Altarelli,
L.~Baudis,
E.~Bellotti, 
R.~Bernabei, 
C.~Bucci,
D. Delepine,
N.~Ferrari,
A.~Incicchitti,
C.~Pobes and A.~Romanino
for discussions and help.
F.F. acknowledges partial finantial support of the European Programs HPRN-CT-2000-00148 and
HPRN-CT-2000-00149.
We acknowledge the partial financial support of the European Program
HPRN-CT-2000-00148.

\appendix

\section{An evidence for neutrino-less double beta decay?}

\begin{figure}
$$\hspace{-5mm}
\includegraphics[width=57mm]{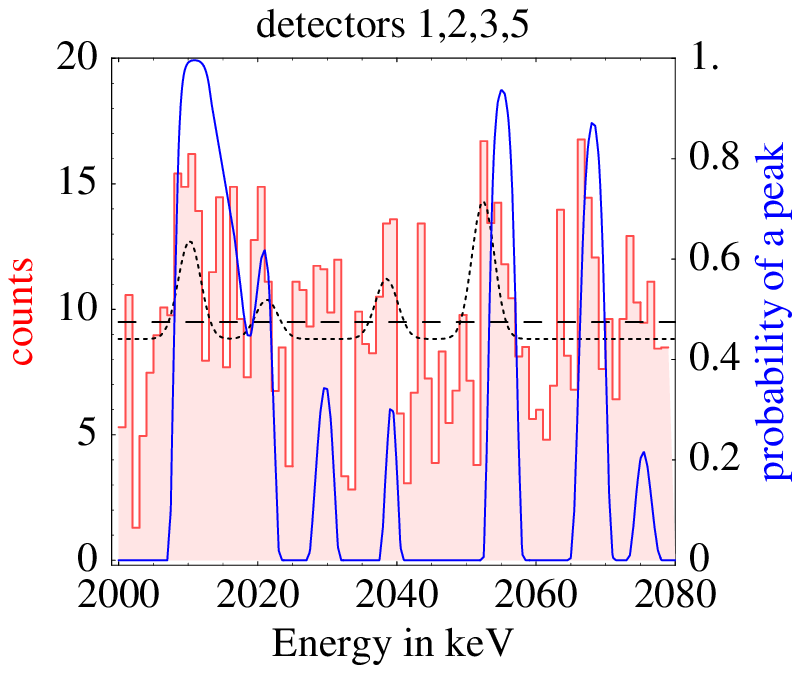}
\includegraphics[width=57mm]{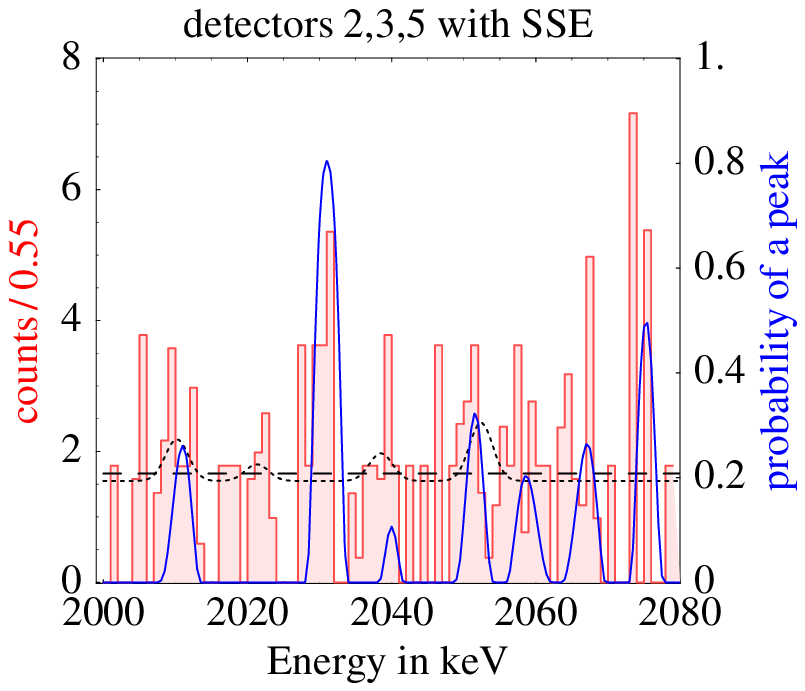}
\includegraphics[width=57mm]{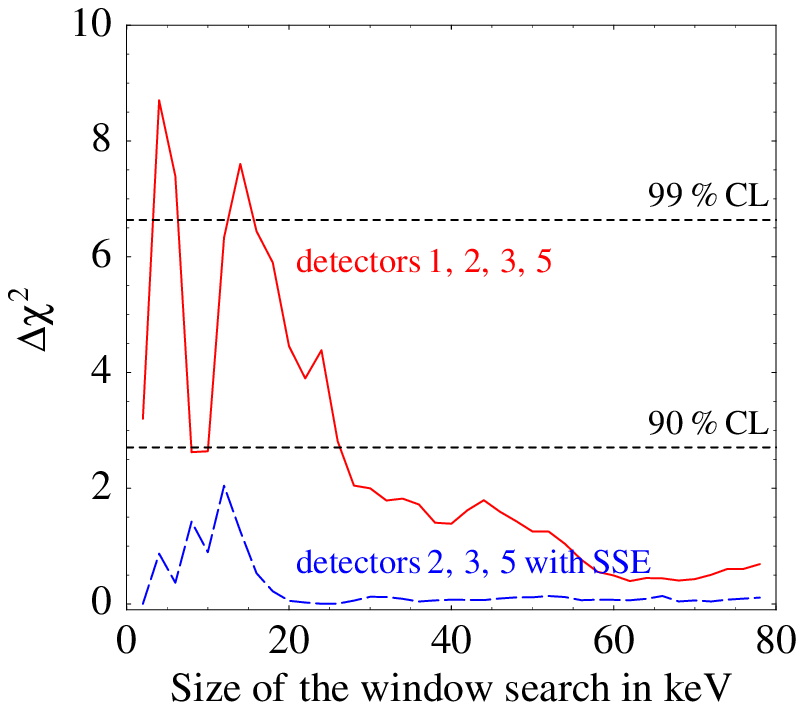}$$
\caption[x]{\em The histograms in the first two figures are two data-sets
(the left ordinates show the scale).
In each case we superimpose the best fit in terms of a constant background (dashed line)
and in terms of a constant background, plus a background due to $\gamma$-peaks of ${}^{214}\hbox{\rm Bi}$,
plus a $0\nu2\beta$ peak at $E = 2039\keV$.
In each plot the continuous blue line shows the likelihood (the right ordinates show the scale)
of having a peak at energy $E$.
The third plot shows how the evidence for a $0\nu2\beta$ signal varies
if, following~\cite{evid}, one fits the data in terms of $0\nu2\beta$ peak plus a constant background,
using only data in a restricted window.
\label{fig:lik}}
\end{figure}

A recent paper~\cite{evid}, which reanalysed 
data of the Heidelberg--Moscow 
experiment at Gran Sasso (HM), 
claimed an evidence for $0\nu 2\beta$ at the $2.2\sigma$ or $3.1\sigma$ level,
when Bayesian or frequentistic techniques are adopted.
The measured value would be $m_{ee}/h = (0.39\pm 0.11)\eV$,
implying almost degenerate neutrinos with mass $m_{\nu_e}\ge |m_{ee}|$,
partly testable at {\sc Katrin}~\cite{katrin}.
Such a result would be of the highest importance.
However, we believe that the published 
data only contain a $\sim 1.5\sigma$ hint
(or less, depending on the data-set used)
 and we would like to explain why in this appendix. 

\medskip

The data in fig.s 2 and 3  of~\cite{evid} are reported
in our fig.s\fig{lik}a and b, where we show
the two histograms with the energy distribution of the 
second data-set (from detectors 1, 2, 3, 5)
and of the third data-set (from detectors 2, 3, 5, with pulse-shape
 discrimination applied in order to
reduce the background). A $0\nu2\beta$ signal would show up as a peak centered at the energy $E =
Q_0=2039.0\keV$ and width equal to the energy resolution,
$\sigma_E =1.59\pm 0.19\keV$.
A quick look at the data shows no clear peak;
so one wonders what is known a priori on the spectrum,
and which is the procedure of analysis to be applied.

Our first step is to repeat the exercise done in~\cite{evid}: we compute
 the likelihood of having one
peak of width $\sigma_E$ centered at a generic energy $E$,
on the top of 
a  constant background of {\em unknown}
intensity $b$ in the energy range $(2000\div 2080)\keV$.
The continuous blue lines in fig.s\fig{lik}a,b (the right ordinates show their scale)
show our results.\footnote{What we
precisely plot is
$1-{\cal L}(\hbox{background})/{\cal L}(\hbox{background}+\mbox{peak at } E)$,
where ${\cal L}(\hbox{background})$ is the likelihood of having a constant background,
while ${\cal L}(\hbox{background}+\mbox{peak at } E)$ is the likelihood of having a constant background plus
a peak at an energy $E$. 
The level of the background and the height of the peak are treated as free parameters.
The likelihoods ${\cal L} \equiv e^{-\chi^2/2}$ are computed as
$\chi^2 = (\sigma_E - 1.59\keV)^2/(0.19\keV)^2 + \sum_i \chi^2_i$,
where the $\chi^2_i$
relative to the $i$th bin (which has 
$n_i$ observed and $\mu_i$ expected  events)
is given by Poissionan statistics as $\chi^2_i = 2 (\mu_i -  n_i \ln \mu_i)$.
Finally, the $\chi^2$ is minimized with respect to the nuisance parameters $b$ and $\sigma_E$.
In Gaussian approximation,
the Bayesian and frequentistic techniques employed in~\cite{evid} both reduce to
this procedure.
We repeated all analyses described in this appendix 
using a Bayesian procedure, finding very little
differences with respect to its Gaussian limit.}
They reasonably agree with the corresponding fig.s 5 and 6 of~\cite{evid},
although we find a somewhat lower probability that a peak is present 
at $E \approx Q_0$ than in~\cite{evid}
(because~\cite{evid} fits data with thinner
binning than in the published data).
In both cases more pronounced peaks seem present at {\em other} energies.

The evidence of~\cite{evid} is obtained by {\em restricting the search window}
to the energy range $E = Q_0 \pm \hbox{few}\cdot\sigma_E$,
such that it does not include the other peaks,
and by fitting the data under the 
hypothesis of a peak at $Q_0$ plus a constant background.
An obvious criticism to this procedure 
is that the data look like a peak over a 
constant background because one 
has chosen the specific window where this happens.
This procedure would be justified if
the size of the search window
were not a significant arbitrary choice,
e.g.\ if one could clearly see in the data
a peak at $Q_0$ emerging over the background.
However, this structure is not clearly visible in the data-sets.
In order to state quantitatively our concern,
we show in fig.\fig{lik}c how the evidence for a $0\nu2\beta$ signal 
(quantified as $\Delta\chi^2 = \chi^2_{{\rm no~}0\nu2\beta}-\chi^2_{\rm best}$)
fluctuates
when one chooses windows of different sizes around $Q_0$.
A `too small' window does not allow to discriminate
the signal from the background, since it can only
be distinguished from the signal exploiting their different
energy dependence.
With an appropriate window the evidence for a peak at $Q_0$ can reach
the $3\sigma$ level. 
Both the evidence and the central value of the signal
change when the size of the window is varied. 
There is almost no evidence
when a large window is chosen; but
if the background were constant, a large window
would be the fairest way to estimate its level.

\begin{figure}
$$\includegraphics[width=7cm]{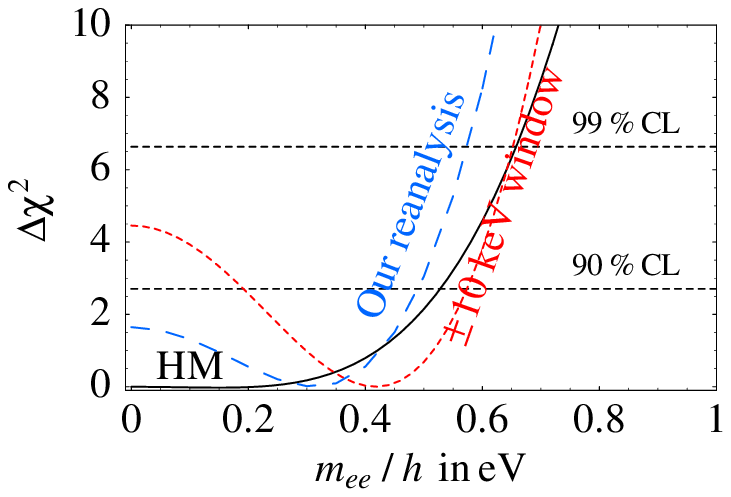}\includegraphics[width=7cm]{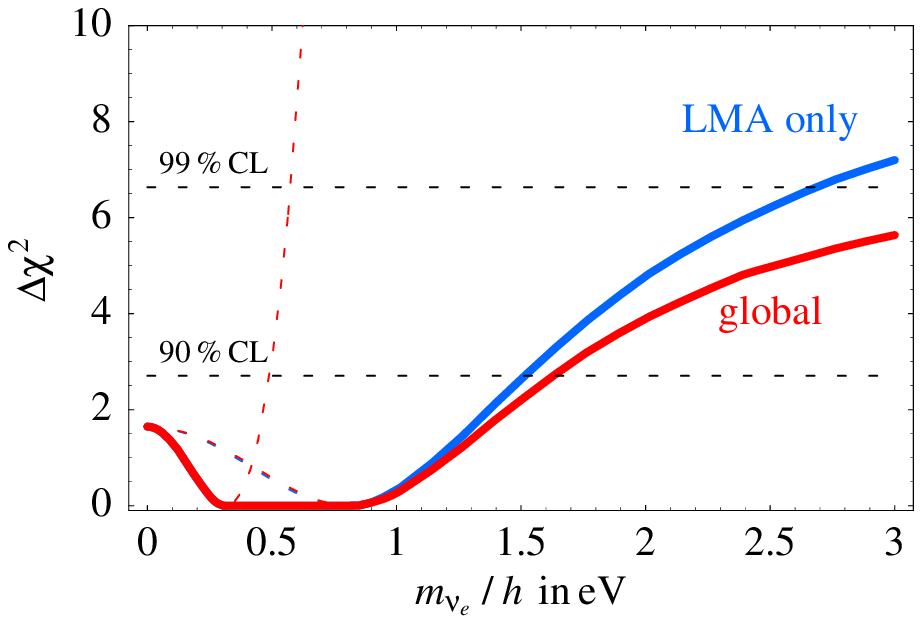}$$
\caption{\em 
Fig.\fig{ultima}a: the $\chi^2(m_{ee}/h)$ extracted in different ways
from the Heidelberg--Moscow data without pulse-shape 
discrimination.
Fig.\fig{ultima}b: bounds on the degenerate neutrino mass,
using our reanalysis of these data.
\label{fig:ultima}}
\end{figure}

In summary: the data contain no evidence for a $0\nu2\beta$ signal
under the assumption of an energy-independent background in the plotted range.
Evidence could only be claimed if the background
had a favourable energy spectrum (smaller around $Q_0$ and larger far from $Q_0$)
and if one could understand it, allowing 
to disentangle the signal from the background.
Therefore the crucial question is:
what is the composition of the background, and in particular
what are the peaks
in the rest of the spectrum, which seem to emerge from
a constant background?
The authors of~\cite{evid} observe that {\em some} peaks 
lie around the known energies of $\gamma$-peaks of ${}^{214}\mbox{Bi}$:
$$E_a = \{2010.7\keV,~2016.7\keV,~2021.8\keV,~2052.9\keV\};$$
this is how they motivate the choice of a small search window that excludes them.

If this is the real composition of the backgrounds,
one should analyse {\em all} data 
assuming that they are composed of:
a constant component, plus the ${}^{214}\mbox{Bi}$ peaks, plus a possible signal $s$.
Therefore we write the number of events as
$$ \frac{dn}{dE} = b  + \sum_a p_a\, \rho(E-E_a) + s\, \rho(E-Q_0)\qquad\hbox{where}\quad
\rho(E) = \frac{e^{-E^2/2\sigma_E^2}}{\sqrt{2\pi}\sigma_E}.$$
The {\em relative} intensity of the ${}^{214}\mbox{Bi}$ $\gamma$-lines is known~\cite{ntab}.
We marginalize the $\chi^2$ with respect to the nuisance parameters
that describe the backgrounds and the energy resolution,
obtaining the $\chi^2$ distribution of the signal, $\chi^2(s)$
(1 degree of  freedom).
In this way we find a weak evidence of a peak at $Q_0$:
$\chi^2(0) - \chi^2_{\rm best}(s)  \approx 2$ (0.5) using the second
(third) data-set; namely, a 1.5$\sigma$ evidence at most.
The best fit is shown in fig.s\fig{lik} (dotted lines).
The constant component of the background is only slightly lower
than under the hypothesis that all events are 
due to a constant background (dashed lines).
In both cases at $Q_0$ there is no peak 
significantly above the background level,
so that we obtain little evidence for a $0\nu2\beta$ signal.
In fig.\fig{ultima}a we show the $\chi^2(m_{ee}/h)$, 
as obtained from three different analyses of the HM 
data without pulse-shape 
discrimination: by choosing
a small search window (as in~\cite{evid}), by modeling the 
background (as proposed here),
and assuming a constant background (as in the HM paper~\cite{HM},
where a slightly different data-set was analysed).
In fig.\fig{ultima}b 
we show the bounds on the degenerate neutrino mass,
using our reanalysis of these data.
Note that the upper bound on $m_{\nu_e}$ in fig.\fig{fit} is stronger, because we used
the HM analysis of data with pulse-shape discrimination.
The connection of massive neutrinos with cosmology is well known (e.g.~\cite{cosm},
for earlier works about neutrinos as warm dark matter see~\cite{libro}).

With actual data and
simulations of the apparatus, it could be possible to do better 
than what we can do, by calculating also the  absolute
intensity of the faint ${}^{214}\mbox{Bi}$ lines in the $(2000\div2080) \keV$ region
based  on other, much more intense lines of this isotope.
According to fig.s~2,3 of~\cite{HM},
the measured intensity of the line at $E\approx 1764.5$ keV is
$25\,\hbox{counts}/\hbox{kg}\cdot\hbox{yr}$.\footnote{Fig.~1 of~\cite{HM}
gives an intensity of $4\,\hbox{counts}/\hbox{kg}\cdot\hbox{yr}$,
6 times smaller.
The reason of the disagreement is that in fig.~1
the counts of a single crystal were divided by total exposure (5 crystals),
 unlike what written in its caption~\cite{KK}.}
Taking into account that 
there is no significant variation of the efficiency
between $1764\keV$ and $Q_0$~\cite{comment},
the line at $1764.5$ keV should have an intensity
of $90\pm 14 ~(25\pm 14)\,\hbox{counts}/\hbox{kg}\cdot\hbox{yr}$,
according to our fit of faint lines
in the second (third) dataset.
Including this extra information in the fit reduces the
significance of the `evidence' to $1\sigma$ ($0.5\sigma$)
in the second (third) dataset.

A significant $0\nu2\beta$ signal could finally result if 
most peaks are not statistical fluctuations, and could be identified.
However a comparison between the two data-sets
discourages such hopes.
We expect that the pulse-shape discrimination should 
suppress the background from $\gamma$-lines 
(since the annihilation energy of the positron may be 
deposited in a relatively wider region) 
more than the hypothetical $0\nu2\beta$ signal
(since the two $e^-$ deposit their energy in a narrow region).
A detailed simulation of the detector is needed
to estimate this issue quantitatively.
The problem is that
fig.s\fig{lik}a,b and our best-fit values indicate
rather that pulse-shape discrimination gives 
an almost equal suppression:
the constant background gets reduced by a factor $\approx 3.5 /\eta$,
the identified $\gamma$-lines by $\approx 3.8 /\eta$, 
the hypothetical peak at $Q_0$ by $\approx 3.2 /\eta$
($\eta=0.55\pm0.10$ is the efficiency of pulse-shape discrimination).
In any case, the assumption that the background in the vicinity of $Q_0$
is particularly small is too important to be just assumed, as done
in~\cite{evid}, and should be demonstrated in some manner.

In conclusion, we do not see a really significant evidence for $0\nu 2\beta$
in the published data.
A better understanding and control of the background
is necessary to learn more from present data (and maybe recognize a signal)
and also in view of future search 
of $0\nu 2\beta$ with ${}^{76}\mbox{Ge}$ detectors.

\frenchspacing
\footnotesize
\begin{multicols}{2}

\end{multicols}
\normalsize

\def\baselinestretch{1}

\newpage

\section*{Addendum A: first KamLAND results}\label{in}
\setcounter{equation}{0}
\renewcommand{\theequation}{\arabic{equation}A}

\begin{figure}
$$\hspace{-8mm}
\includegraphics[width=8.5cm]{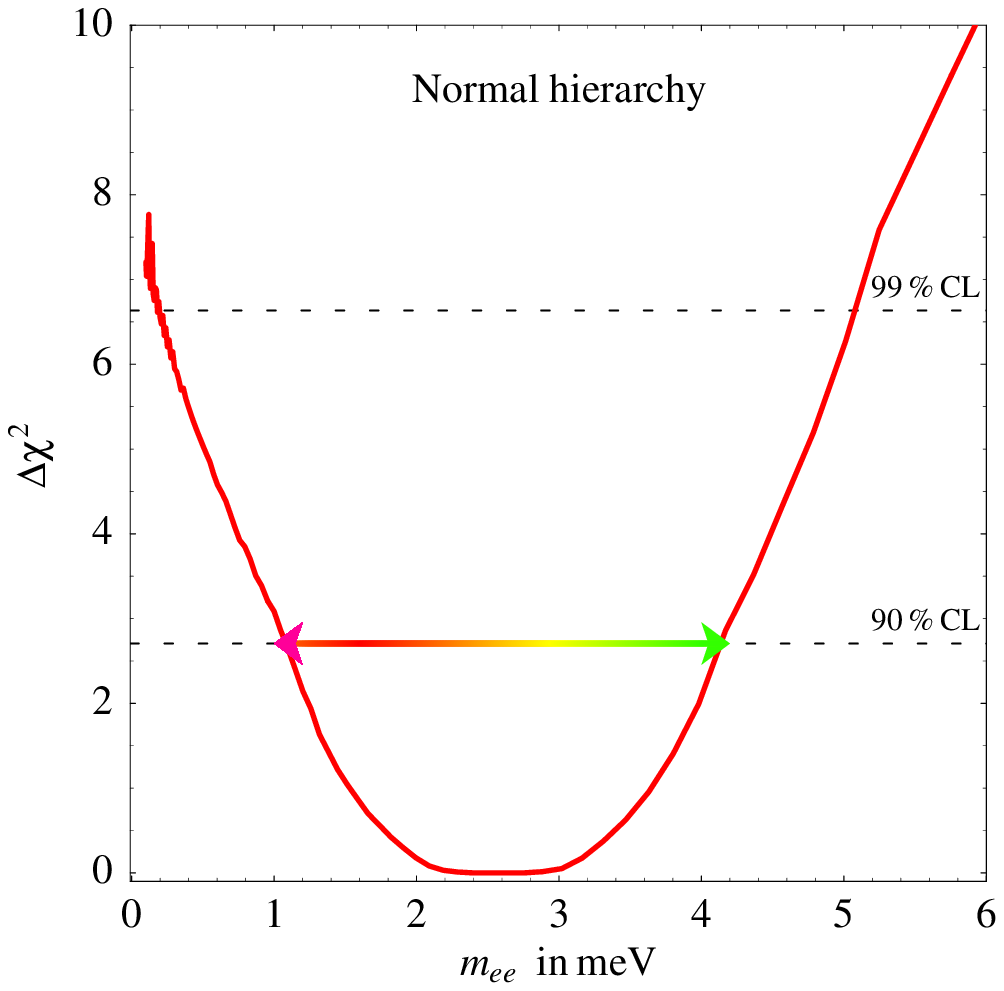}\hspace{1cm}\includegraphics[width=8.5cm]{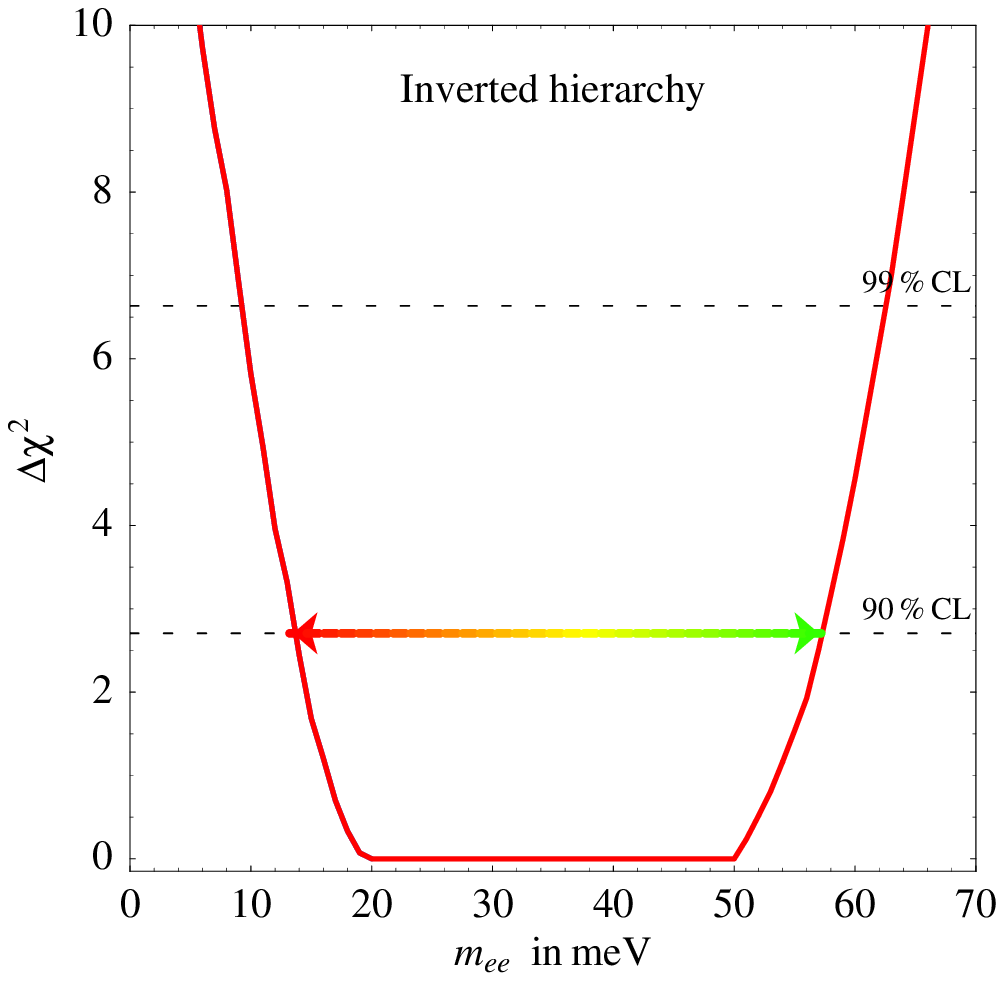}$$
$$\hspace{-8mm}
\includegraphics[width=8.5cm]{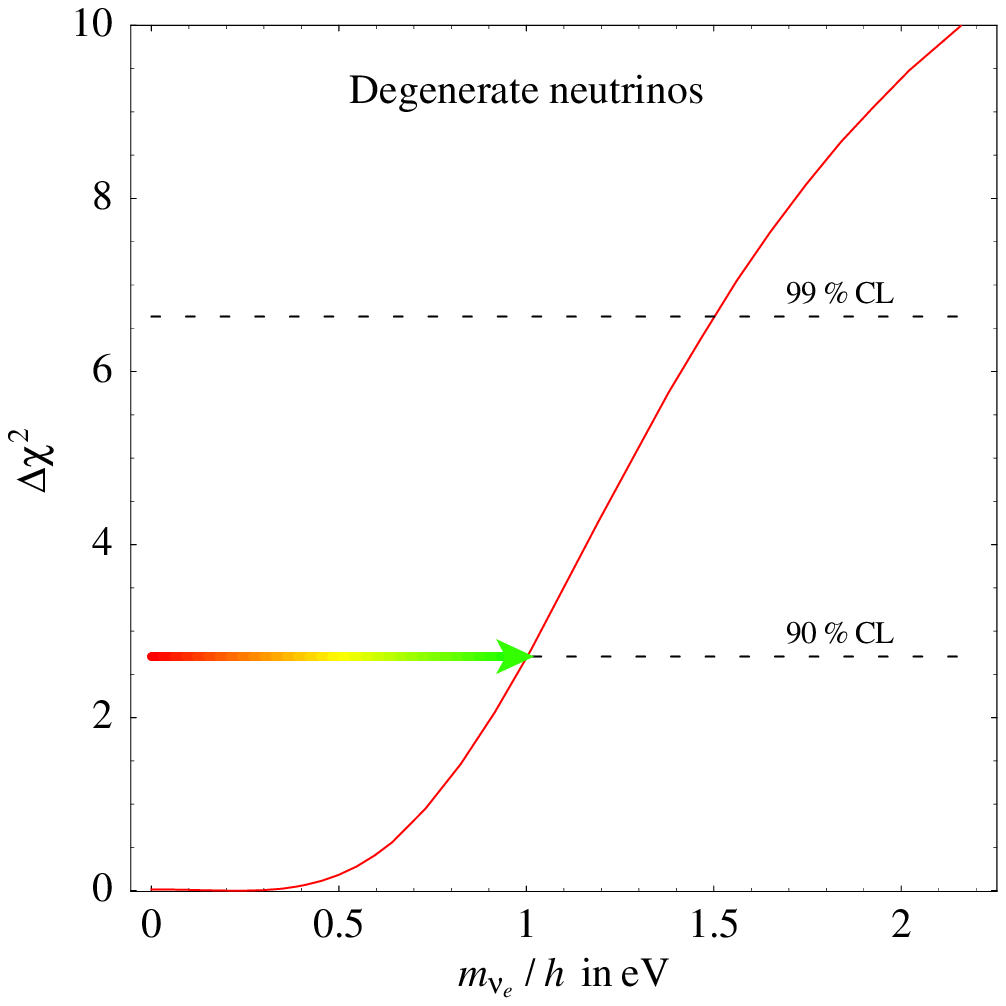}\hspace{1cm}\includegraphics[width=8.5cm]{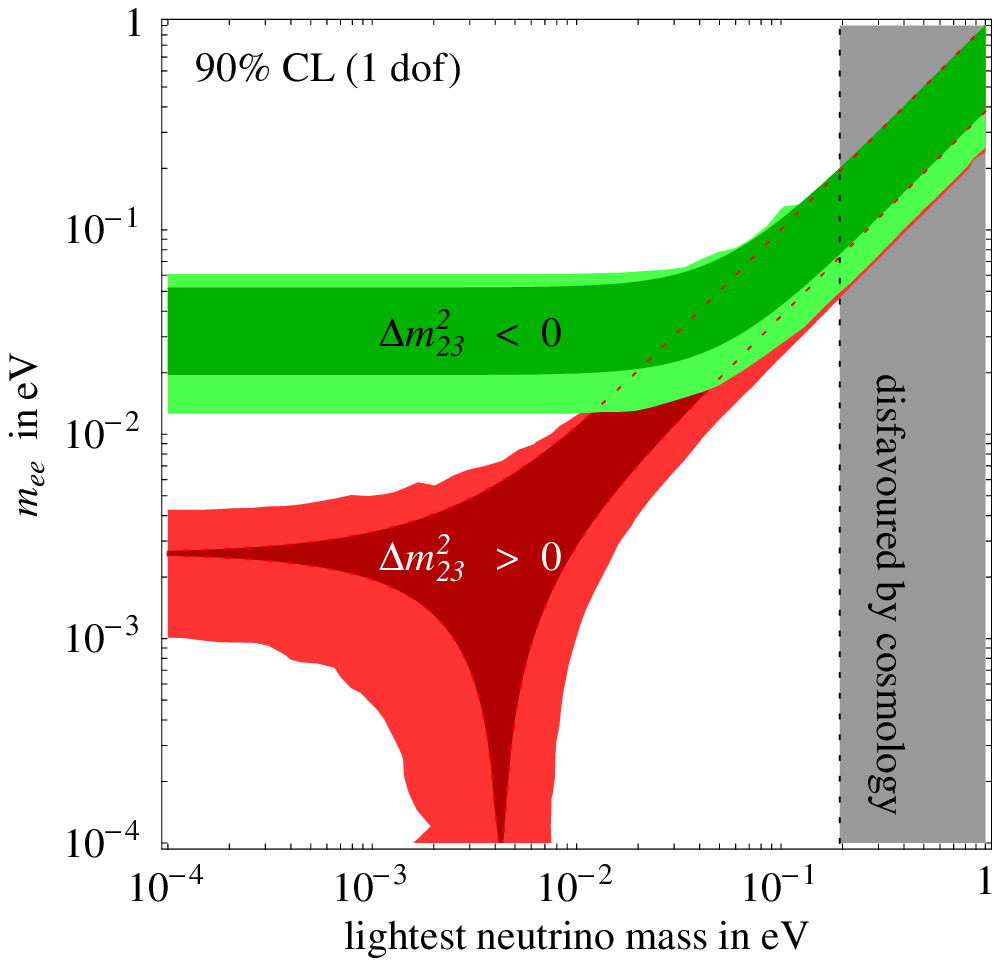}$$

\caption{\label{fig:postKL}\em Predictions for $|m_{ee}|$ assuming a hierarchical 
(fig.\fig{postKL}a) and
inverted (fig.\fig{postKL}b)  neutrino spectrum. 
In fig.\fig{postKL}c we update the upper bound on the mass of quasi-degenerate
neutrinos implied by $0\nu2\beta$ searches.
The factor $h\approx 1$ parameterizes the uncertainty in the 
nuclear matrix element (see
sect.\ \protect{\ref{sect:noOscExps}}).
In fig.\fig{postKL}d we plot the $90\%$ CL range for $m_{ee}$
as function of the lightest neutrino mass,
thereby covering all spectra.
The darker regions show how the $m_{ee}$ range would shrink if
the present best-fit values of oscillation parameters were confirmed with negligible error.
}\end{figure}

In this addendum we  update our previous results by including
new recent data.
Concerning solar data, we include the KamLAND data~\cite{KamLAND}
that confirm the LMA solution,
the most recent data from GNO and the full set of spectral and day/night SNO data~\cite{SNOlast}.
Concerning atmospheric data, we include the latest SK and K2K data~\cite{SK2K}.
In view of these improvements, our knowledge of neutrino oscillation parameters
can now be approximatively summarized as
$$\Delta m^2_{12} = (7.1 \pm 0.6)10^{-5}\eV^2,\qquad
|\Delta m^2_{23}| = (2.7\pm 0.4) 10^{-3}\eV^2,$$
$$\tan^2\theta_{12} = 0.45 \pm  0.06,\qquad
\sin^22\theta_{23} = 1.00 \pm 0.04,\qquad
\sin^2 2 \theta_{13}  = 0\pm 0.065. $$
Our inferences on the parameter $|m_{ee}|$ probed by $0\nu2\beta$ experiments
are obtained by marginalizing the full joint probability distribution
for the oscillation parameters.
We employ the fit of solar and KamLAND data presented in~\cite{oursunfit}
(6 out of 8 other similar analyses~\cite{lastsunfit} agree with~\cite{oursunfit}).
Our updated results, shown in fig.\fig{postKL}a,b,c,d, are similar to
the lines ``LMA only'' in the corresponding figures\fig{dir},\fig{inva},\fig{fit} and\fig{summary} 
in the original version of this work.
Assuming three CPT-invariant massive Majorana neutrinos we get
\begin{itemize}

\item[a)]  In the case of {\bf normal hierarchy} (i.e.\ $m_1\ll m_2\ll m_3$, or $\Delta m^2_{23}>0$)
the $ee$ element of the neutrino mass matrix probed by
$0\nu2\beta$ decay experiments
can be written as $|m_{ee}| = |e^{2i \alpha}  m_{ee}^{\rm sun} + e^{2i\beta} m_{ee}^{\rm atm}|$
(see section~\ref{sect:d})
where $\alpha,\beta$ are unknown Majorana phases and 
the `solar' and `atmospheric' contributions can be predicted from oscillation data.
Including KamLAND data, that restrict the LMA range  of $\Delta m^2_{12}$, 
the solar contribution to $m_{ee}$
is $m_{ee}^{\rm sun} =( 2.6\pm 0.4)~\,\hbox{meV}$.
The bound on $\theta_{13}$ from CHOOZ, SK and K2K implies $m_{ee}^{\rm atm} < 2\,\hbox{meV}$ at 99\% CL.
By combining these two contributions we get
the range
\begin{equation}
\label{eq:neqh}
|m_{ee}|=(1.1\div 4.1) \meV\hbox{ at $90\%$ CL}\qquad\hbox{ and }\qquad
|m_{ee}|=(0.2\div 5.1) \meV \hbox{ at $99\%$ CL.}
\end{equation}
The precise prediction is shown in fig.\fig{postKL}a.

\item[b)] In the case of {\bf inverted hierarchy} (i.e.\ $m_3\ll m_1\approx m_2$, or $\Delta m^2_{23}<0$), from
the prediction
$|m_{ee}| \approx (\Delta m^2_{23})^{1/2}\times
|\cos^2\theta_{12} +  e^{2i\alpha} \sin^2 \theta_{12}|\cos^2 \theta_{13}$
(see section~\ref{sect:i}) 
we get the range
\begin{equation}
\label{eq:newi}
|m_{ee}|=(14\div 57) \meV\hbox{ at $90\%$ CL}\qquad\hbox{ and }\qquad
|m_{ee}|=(9.2 \div 61) \meV \hbox{ at $99\%$ CL.}
\end{equation}The precise prediction is shown in fig.\fig{postKL}b.
The inclusion of KamLAND data does not significantly restrict the $|m_{ee}|$ {\em range} because
the main uncertainty is due to the Majorana phase $\alpha$
not to the oscillation  parameters.
If the present central values were confirmed with infinite precision we would still have the loose range
$|m_{ee}|=(20\div 52) \meV$ at any CL.
The uncertainty in the lower (upper) bound is today dominated by the error on $\theta_{12}$ ($\Delta m^2_{23}$).
\end{itemize}
The $|m_{ee}|$ ranges for normal and inverted hierarchy no longer overlap. 
Values of $|m_{ee}|$ outside these ranges are possible if the lightest neutrino mass
is not negligible, as shown in fig.\fig{postKL}d.
If $\Delta m^2_{23}<0$ a non zero lightest neutrino mass does not allow
smaller $|m_{ee}|$ than in case b): eq.\eq{newi} is a true lower bound.
The darker regions in fig.\fig{postKL}d show how the predicted range of $|m_{ee}|$ would shrink 
if the present best-fit values of
oscillation  parameters were confirmed with infinite precision.
The two funnels present for $\Delta m^2_{23}>0$
have an infinitesimal width because we assume $\theta_{13} = 0$
and would have a finite width if $\theta_{13}\neq 0$.


\begin{itemize}

\item[c)]
The bound on the mass $m_{\nu_e}$ of {\bf almost-degenerate} neutrinos
implied by present $0\nu 2\beta$ data is 
\begin{equation}
\label{eq:newd}
m_{\nu_e}< 1.0 ~(1.5)\, h  \eV\qquad\hbox{ at 90 ($99\%$) CL.}
\end{equation}
The precise bound is shown in fig.\fig{postKL}c.
The factor $h\approx 1$ parameterizes the uncertainty in the 
nuclear matrix element (see
sect.\ \protect{\ref{sect:noOscExps}}).
As discussed in section~\ref{sect:d} (see also~\cite{minakata}) this bound is based on the fact that
a large $m_{\nu_e}$ is compatible with the upper bound on $|m_{ee}|$ 
only if  $\theta_{12}\approx \pi/4$.
Maximal solar mixing was already disfavoured at almost $3\sigma$ before KamLAND
(fig.\fig{fit}) and is now disfavoured at almost $4\sigma$ (fig.\fig{postKL}c).
KamLAND alone does not yet disfavour $\theta_{12}\approx \pi/4$.
The bound on neutrino masses in eq.\eq{newd},  obtained combining
oscillation  and $0\nu2\beta$ data under the assumption that neutrinos are Majorana particles,
can be directly compared 
with the bound from direct searches of neutrino mass in tritium decays,
$m_{\nu_e}  <2.2\ \mbox{eV}$ at $95\%$ CL.
We employ the bound on $|m_{ee}|$ from
 Heidelberg--Moscow (HM) data, as analyzed by the collaboration.
\end{itemize}
A reanalysis~\cite{evid2} of HM data~\cite{HM2} claimed evidence
 for $0\nu2\beta$.
Since this is an important but controversial issue, even minor points have been carefully 
debated~\cite{Aalseth2, Zde,Klapdor2, debate,Ianni}.
We here summarize what we believe is the main point.
The reanalysis in~\cite{evid2} was restricted to the particular subset of data
that seems to contain an evidence, neglecting the rest.
Reanalysing all data using what we know about the signal and about the background we found
a $1.5\sigma$ hint at most (see appendix A of this paper\footnote{The authors of~\cite{evid2}
claim that the restriction in the window search is not a critical arbitrary choice,
in apparent disagreement with our appendix.
This claim refers to a fit of an average sample of {\em simulated} data
(generated under some assumption).
Our claim is that the restriction in the window search  is a critical arbitrary choice 
  when analyzing the {\em real} data: this can be seen from our fig.\fig{lik}c.}).
Other reanalyses 
performed along these lines and precisely described by their authors find
$1.1\sigma$~\cite{Ianni}, $1.46\sigma$~\cite{Klapdor2},
less than $1\sigma$~\cite{Zde} (these authors combine
HM~\cite{HM2} and IGEX~\cite{IGEX2} data.
Both experiments use $^{76}$Ge with a similar energy resolution and background level.
IGEX has about 5 times less statistic and finds a slight 
  deficit of events around the $0\nu2\beta$ $Q$ value, where HM   
  finds a slight excess).
In conclusion, the hint for $0\nu 2\beta$ is not statistically significant.



\footnotesize

\begin{multicols}{2}

\end{multicols}
\label{out}

\def\baselinestretch{1.05}

\newpage

\label{insalt}

\normalsize
\section*{Addendum B: `salt' SNO data}
\setcounter{equation}{0}
\renewcommand{\theequation}{\arabic{equation}B}

\begin{figure}
$$\hspace{-8mm}
\includegraphics[width=8.5cm]{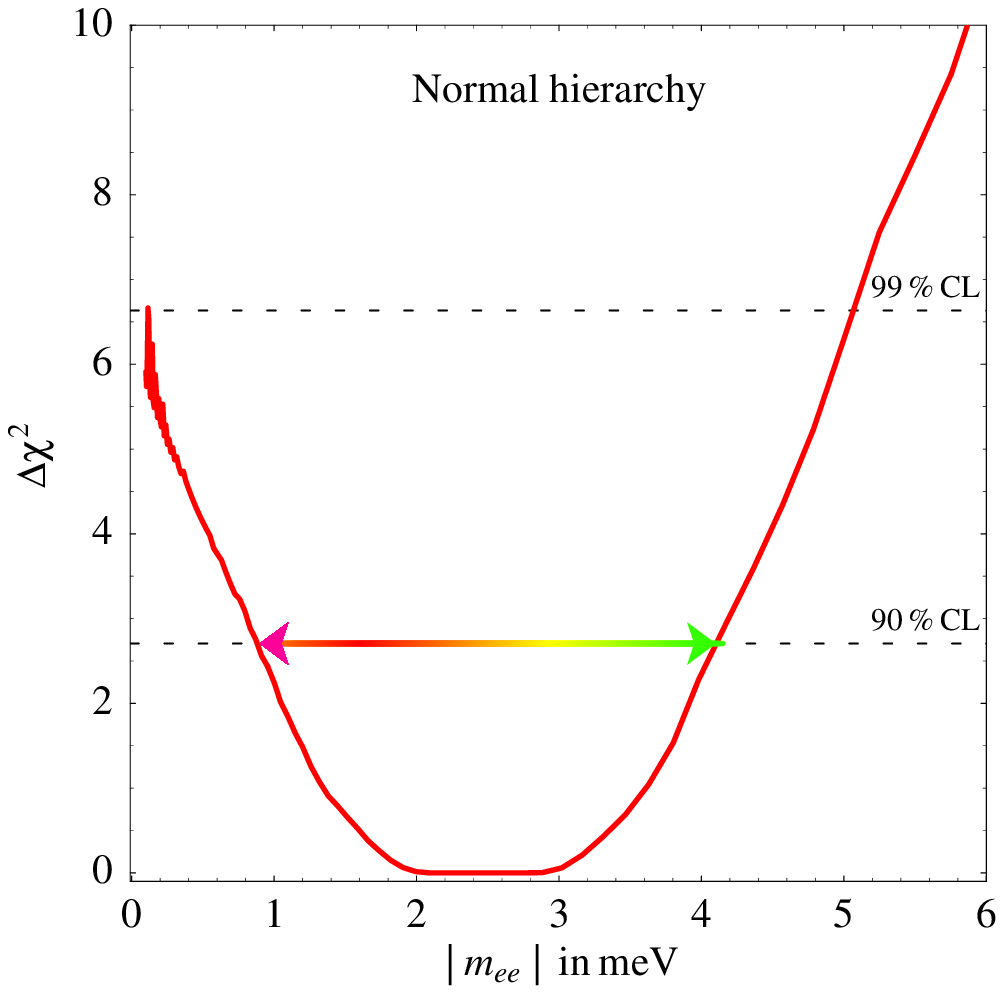}\hspace{1cm}\includegraphics[width=8.5cm]{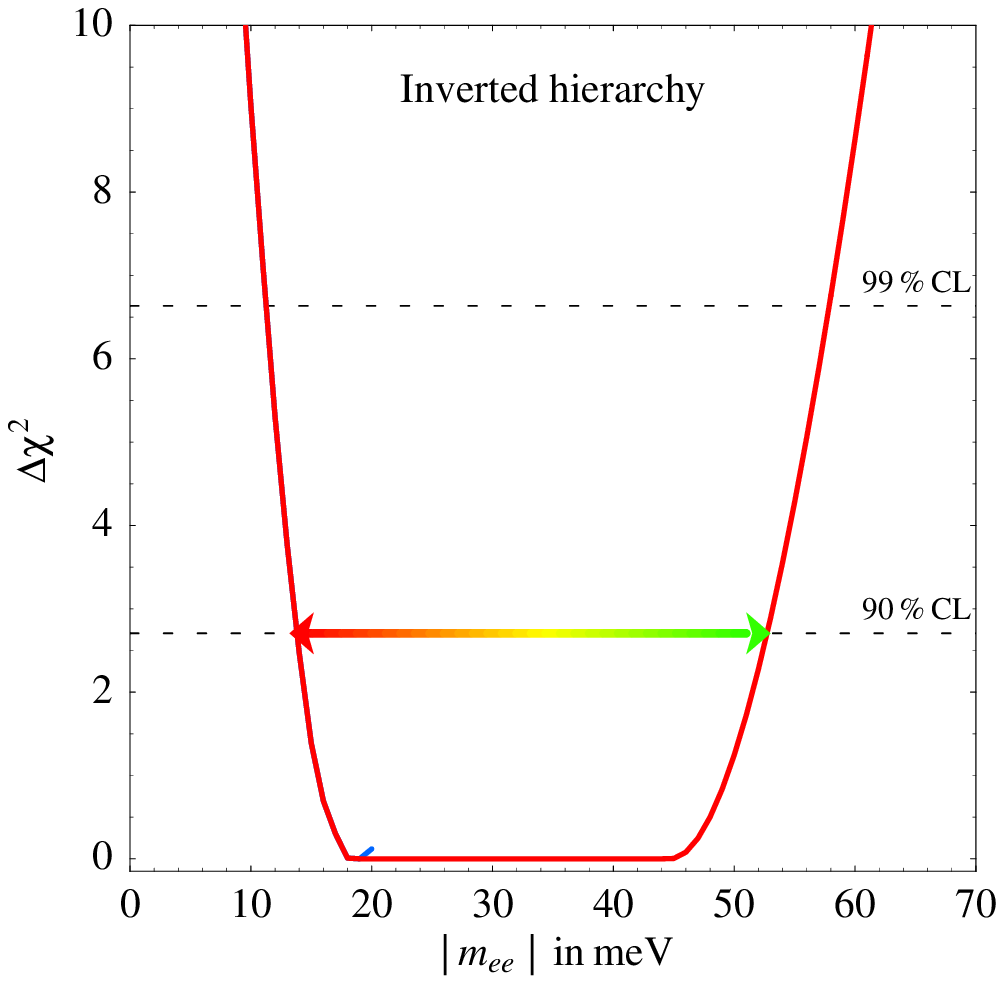}$$
$$\hspace{-8mm}
\includegraphics[width=8.5cm]{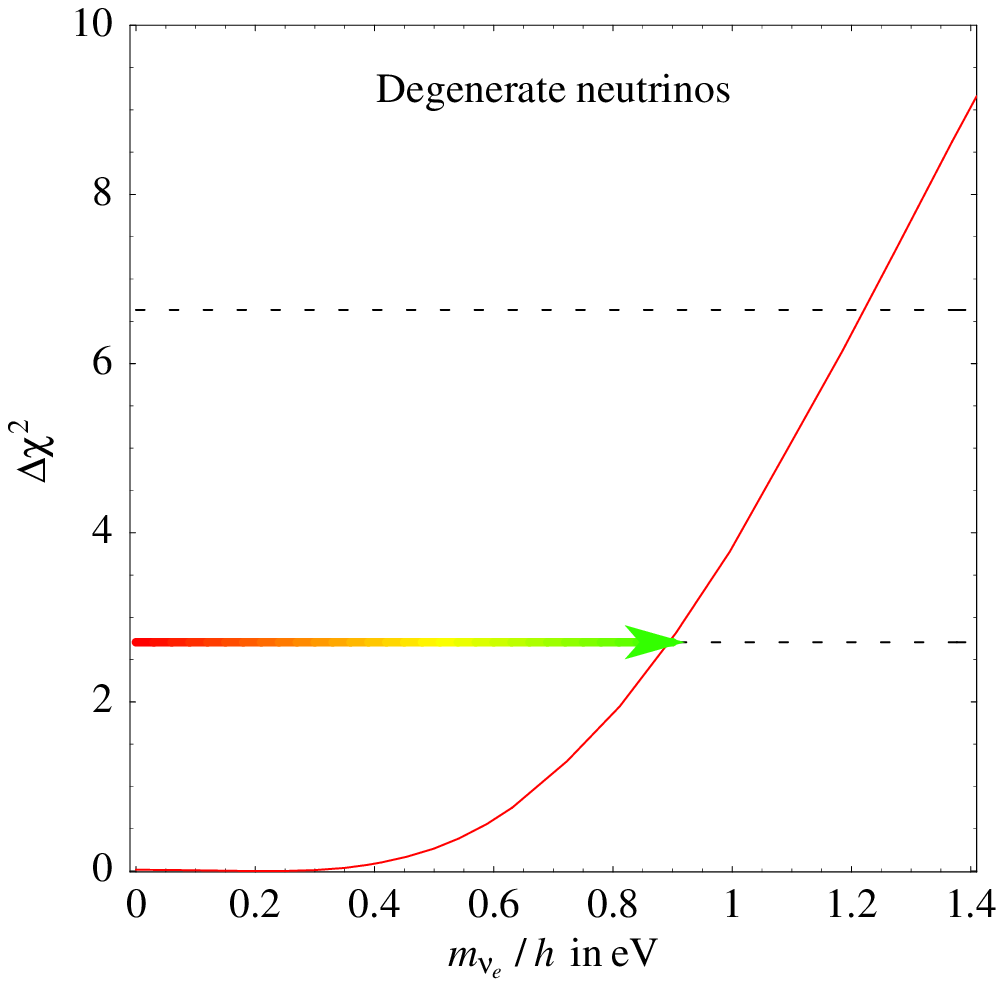}\hspace{1cm}\includegraphics[width=8.5cm]{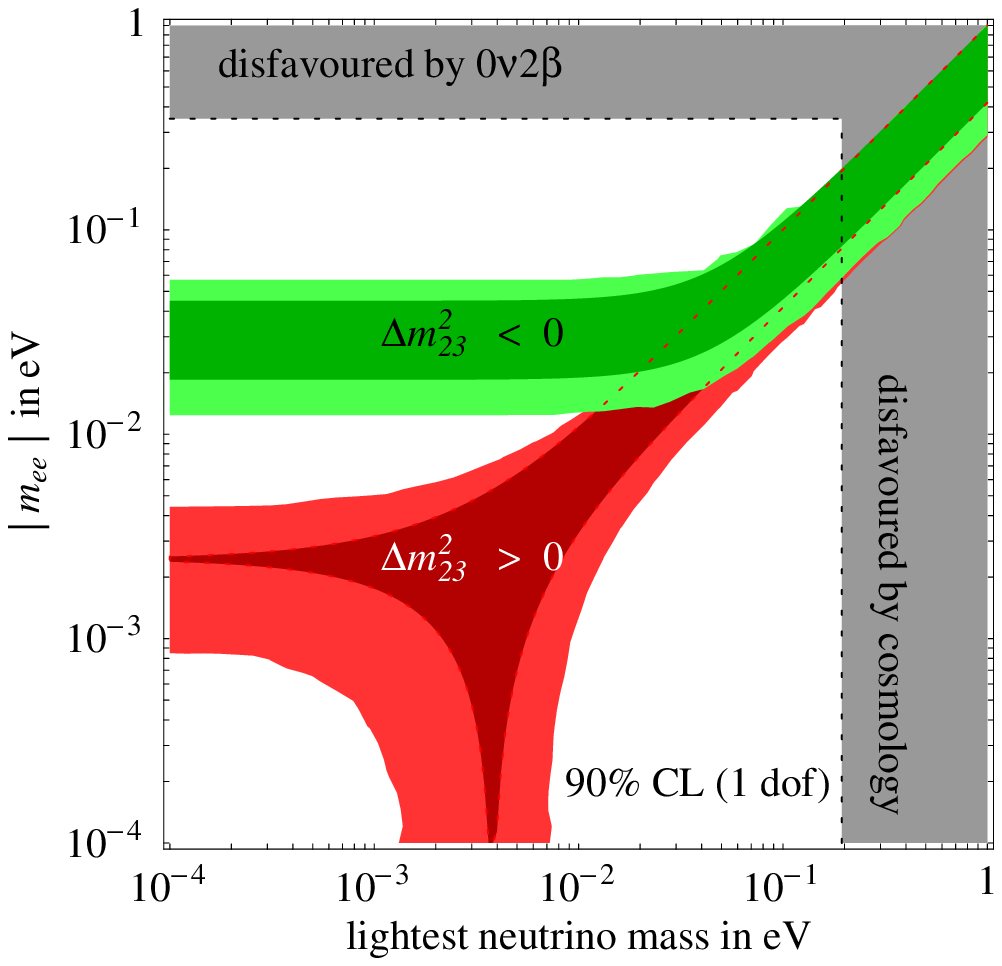}$$

\caption{\label{fig:postsalt}\em Predictions for $|m_{ee}|$ assuming a hierarchical 
(fig.\fig{postsalt}a) and
inverted (fig.\fig{postsalt}b)  neutrino spectrum. 
In fig.\fig{postsalt}c we update the upper bound on the mass of quasi-degenerate
neutrinos implied by $0\nu2\beta$ searches.
The factor $h\approx 1$ parameterizes the uncertainty in the 
nuclear matrix element (see sect.\ 2.1).
In fig.\fig{postsalt}d we plot the $90\%$ CL range for $m_{ee}$
as function of the lightest neutrino mass,
thereby covering all spectra.
The darker regions show how the $m_{ee}$ range would shrink if
the present best-fit values of oscillation parameters were confirmed with negligible error.
}\end{figure}

We update our previous 
results by including `salt' SNO data and
the most recent data from GNO and SAGE~\cite{SNOsalt}.
Concerning atmospheric data, we take into account the latest 
Super-Kamiokande (SK) and K2K analyses~\cite{SK2K03}, which
reduced the best fit value of the atmospheric mass splitting.
In view of these improvements, our knowledge of neutrino 
oscillation parameters
can now be approximatively summarized as\footnote{To give a feeling of
how accuratelty present fits can be summarized in this way, 
we dicuss the case of the solar mixing angle.
According to our analysis, the $1\sigma$ ranges around the best-fit value are
$\tan^2\theta = 0.414^{+0.05}_{-0.04}$.
Since errors are more asymmetric at larger confindence levels
we prefer to report the symmetric $1\sigma$ range that gives the
precise $99\%$ CL range.}

$$\Delta m^2_{12} = (7.2 \pm 0.7)10^{-5}\eV^2,\qquad
|\Delta m^2_{23}| = (2.0\pm 0.4) 10^{-3}\eV^2,$$
$$\tan^2\theta_{12} = 0.44 \pm  0.05,\qquad
\sin^22\theta_{23} = 1.00 \pm 0.04,\qquad
\sin^2 2 \theta_{13}  = 0\pm 0.085. $$
Our inferences on the parameter $|m_{ee}|$ probed by $0\nu2\beta$ experiments
are obtained by marginalizing the full joint probability distribution
for the oscillation parameters.
We employ the fit of solar and KamLAND data presented in~\cite{sunfitsalt}.
Assuming three CPT-invariant massive Majorana neutrinos we get
\begin{itemize}

\item[a)]  In the case of {\bf normal hierarchy} (i.e.\ $m_1\ll m_2\ll m_3$, or $\Delta m^2_{23}>0$)
the $ee$ element of the neutrino mass matrix probed by
$0\nu2\beta$ decay experiments
can be written as $|m_{ee}| = |e^{2i \alpha}  m_{ee}^{\rm sun} + e^{2i\beta} m_{ee}^{\rm atm}|$
(see section~2.3)
where $\alpha,\beta$ are unknown Majorana phases and 
the `solar' and `atmospheric' contributions can be predicted from oscillation data.
Including KamLAND data, that restrict the LMA range  of $\Delta m^2_{12}$, 
the solar contribution to $m_{ee}$
is $m_{ee}^{\rm sun} =( 2.45\pm 0.23)~\,\hbox{meV}$.

The bound on $\theta_{13}$ from CHOOZ, SK and K2K implies $m_{ee}^{\rm atm} < 2\,\hbox{meV}$ at 99\% CL.
By combining these two contributions we get
the range
\begin{equation}
\label{eq:neqh}
|m_{ee}|=(0.9\div 4.1) \meV\hbox{ at $90\%$ CL}\qquad\hbox{ and }\qquad
|m_{ee}|=(0\div 5.1) \meV \hbox{ at $99\%$ CL.}
\end{equation}
The precise prediction is shown in fig.\fig{postsalt}a.

\item[b)] In the case of {\bf inverted hierarchy} (i.e.\ $m_3\ll m_1\approx m_2$, or $\Delta m^2_{23}<0$), from
the prediction
$|m_{ee}| \approx (\Delta m^2_{23})^{1/2}\times
|\cos^2\theta_{12} +  e^{2i\alpha} \sin^2 \theta_{12}|\cos^2 \theta_{13}$
(see section~2.4) 
we get the range
\begin{equation}
\label{eq:newi}
|m_{ee}|=(14\div 53) \meV\hbox{ at $90\%$ CL}\qquad\hbox{ and }\qquad
|m_{ee}|=(11 \div 58) \meV \hbox{ at $99\%$ CL.}
\end{equation}The precise prediction is shown in fig.\fig{postsalt}b.
The inclusion of SNO `salt' data does not significantly restrict the $|m_{ee}|$ {\em range} because
the main uncertainty is due to the Majorana phase $\alpha$,
no longer due to the oscillation  parameters.
If the present best-fit values were confirmed with infinite precision we would still have the loose range
$|m_{ee}|=(19\div 45) \meV$ at any CL. 
Ref.~\cite{MuraPenya} found a value in agreement with 
our results and emphasized that the lower bound on $|m_{ee}|$
(here shown in the left part of fig.~\ref{fig:postsalt}b)
remains valid even if KamLAND or one
of the solar data is dropped from 
a global analysis (the bound is `robust').

\end{itemize}
The $|m_{ee}|$ ranges for normal and inverted hierarchy do not overlap. 
Values of $|m_{ee}|$ outside these ranges are possible if the lightest neutrino mass
is not negligible, as shown in fig.\fig{postsalt}d.
If $\Delta m^2_{23}<0$ a non zero lightest neutrino mass does not allow
smaller $|m_{ee}|$ than in case b): eq.\eq{newi} is a true lower bound.
The darker regions in fig.\fig{postsalt}d 
show how the predicted range of $|m_{ee}|$ would shrink 
if the present best-fit values of
oscillation  parameters were confirmed with infinite precision.
The two funnels present for $\Delta m^2_{23}>0$
have an infinitesimal width because we assume $\theta_{13} = 0$
and would have a finite width if $\theta_{13}\neq 0$.


\begin{itemize}

\item[c)]
The bound on the mass $m_{\nu_e}$ of {\bf almost-degenerate} neutrinos
implied by present $0\nu 2\beta$ data is 
\begin{equation}
\label{eq:newd}
m_{\nu_e}< 0.9 ~(1.2)\, h  \eV\qquad\hbox{ at 90 ($99\%$) CL.}
\end{equation}
The precise bound is shown in fig.\fig{postsalt}c.
The factor $h\approx 1$ parameterizes the uncertainty in the 
nuclear matrix element (see sect.\ 2.1).
As discussed in section~2.5  this bound is based on the fact that
a large $m_{\nu_e}$ is compatible with the upper bound on $|m_{ee}|$ 
only if  $\theta_{12}\approx \pi/4$.
Maximal solar mixing was already disfavoured at almost $4\sigma$ before salt SNO data
and is now disfavoured at $5.4\sigma$ (fig.\fig{postsalt}c).
We employ the bound on $|m_{ee}|$ from
 Heidelberg--Moscow (HM) data, as analyzed by the collaboration.
\end{itemize}
The bound on neutrino masses in eq.\eq{newd},  obtained combining
oscillation  and $0\nu2\beta$ data under the assumption 
that neutrinos are Majorana particles,
can be directly compared 
with the bound from direct searches 
of neutrino mass in tritium decays,
$m_{\nu_e}  <2.2\ \mbox{eV}$ at $95\%$ CL, or with 
the tight limits obtained in cosmology \cite{WMAP},
$m_{\nu_e}  <0.23\ \mbox{eV}$ at $95\%$ CL. 
Fig.~\ref{fig:postsalt}d also shows bounds from cosmology~\cite{WMAP} 
and from the Heidelberg-Moscow~\cite{HM} $0\nu2\beta$ experiment (assuming $h=1$).
Both bounds are somewhat controversial,
and in both cases hints that some effect has been detected
have been suggested by some 
alternative analyses~\cite{hintHM,hintC}, not reported here.
In practice, these bounds can be 
relaxed by a factor of two in 
more conservative analyses.

For $\beta$ decay the predictions are  (99\% CL range):
$m_{\nu_e} = (3.9\div 10.8)\meV$ for normal hierarchy and
$m_{\nu_e} = (45\pm 4)\meV$ for inverted hierarchy.
This last value is 5 times below the planned sensitivity of Katrin.

\footnotesize

\begin{multicols}{2}

\end{multicols}

\label{outsalt}

\end{document}